\documentclass[letterpaper,english,reprint, aps]{revtex4-1}
\pdfoutput=1
\usepackage[T1]{fontenc}
\usepackage[latin9]{inputenc}
\usepackage{amsmath}
\usepackage{amssymb}
\usepackage{graphicx}
\usepackage{esint}
\usepackage{cleveref}

\makeatletter

\pdfpageheight\paperheight
\pdfpagewidth\paperwidth

 \@ifundefined{textcolor}{}
 {%
   \definecolor{BLACK}{gray}{0}
   \definecolor{WHITE}{gray}{1}
   \definecolor{RED}{rgb}{1,0,0}
   \definecolor{GREEN}{rgb}{0,1,0}
   \definecolor{BLUE}{rgb}{0,0,1}
   \definecolor{CYAN}{cmyk}{1,0,0,0}
   \definecolor{MAGENTA}{cmyk}{0,1,0,0}
   \definecolor{YELLOW}{cmyk}{0,0,1,0}
 }

\makeatother

\usepackage{babel}
\begin{document}

\title{Segment-wise Description of the Dynamics of Traffic Congestion }

\author{T.S. Choi}

\affiliation{Department of Physics, Hong Kong University of Science and Technology,
Hong Kong}

\author{Kiwing To}

\affiliation{Institute of Physics, Academia Sinica, Taipei}

\author{K.Y. Michael Wong}

\affiliation{Department of Physics, Hong Kong University of Science and Technology,
Hong Kong}

\date{\today}
\begin{abstract}
We compare the point-wise and segment-wise descriptions of the traffic
system. Using real data from the Taiwan highway system with a tremendous
volume of segment-wise data, we find that the segment-wise description
is much more informative of the evolution of the system during congestion.
Congestion is characterized by a loopy trajectory in the fundamental
diagram. By considering the area enclosed by the loop, we find that
there are two types of congestion dynamics -{}- moderate flow and
serious congestion. They are different in terms of whether the area
enclosed vanishes. Data extracted from the time delays of individual
vehicles show that the area enclosed is a measure of the economic
loss due to congestion. The use of the loss area in helping to understand
various road characteristics is also explored. 
\end{abstract}
\maketitle

\section{\label{sec:level1}Introduction}

With the rapid development all over the world in the past few centuries,
there is a huge demand for logistics or long distance transportation
through the highway system. According to statistics from the Environmental
Protection Administration Executive Yuan in Taiwan, the number of
registered vehicles increased from $6.8$ million in $2010$ to $7.9$
million in $2017$. In Hong Kong, the number of registered vehicles
is also rising continuously, from $0.57$ million in $2008$ to $0.77$
million in $2018$. This rising demand cannot be satisfied by the
construction of new highways. Congestion is inevitably more frequent
and more serious. Journeys of drivers are delayed whenever rush hours
come. The increase of traveling time indicates an economic loss to
the society. It is important to understand congestion so as to minimize
the loss. Research was conducted on modelling and data analysis, in
order to prevent the traffic system from entering the congestion phase
\citep{daganzo2007urban,cole2003pricing,yang2011managing}. However,
the dynamics during congestion is not well-studied and still under
debate \citep{schonhof2009criticism,kerner2016failure,kerner2014probabilistic,schonhof2007empirical}.

Conventional traffic models use observables at a point to describe
the traffic condition \citep{lighthill1955kinematic,nagel1992cellular,kerner1996experimental}.
For instance, if the speed detected is low, the traffic is considered
to be congested. However, since stop-and-go waves exist in the traffic
system during congestion, the speed at a point would fluctuate rapidly.
The point-wise description is not capable of determining the traffic
condition in this case. This problem can be solved by using a segment-wise
description, as will be verified through simulations in this paper.

This study is based on real data from the Taiwan highway system. The
data collection system was set up for the purpose of toll collection.
Hence the data consists of the time information of individual vehicles
passing through the sensors along their highway journeys. Highway
segments can then be demarcated by successive sensors, enabling us
to extract a tremendous volume of segment-wise information.

With the more precise segment-wise description of traffic congestions,
we are able to trace the evolution of the traffic system during congestion,
obtaining much clearer trajectories in the graph of flux versus
density, which is also known as the fundamental diagram of traffic. We
find that there are two different dynamics during congestion, the
moderate flow and the serious congestion. They can be distinguished
by the area enclosed by their trajectories in the fundamental diagram.

There is a further advantage of the individualized vehicle data from
the toll collection system. In the past, the traffic data was mainly
flux and averaged speed recorded at a point \citep{kerner1996experimental,sugiyama2008traffic}.
The information of vehicle identity was missing. The delay suffered
by a vehicle across two points on highway could not be measured. From
the individual trajectory data from the toll collection system, we
can compute the delay suffered by each vehicle, and hence deduce the
economic loss incurred during a congestion event. We also show that
the area enclosed in the fundamental diagram during congestion reflects
the economic loss incurred.

The article is organized as follows. In Sec. \ref{sec:Description-of-Traffic},
we briefly discuss the conventional models in traffic theory and the
debates of behaviors in congestion. We then introduce the point-wise
and segment-wise descriptions. We illustrate the insufficiency of
the point-wise description through the simulations of the optimal
velocity model, and how it was outperformed by the segment-wise description.
In Sec. \ref{sec:Congestion-in-Reality}, behaviors of real traffic
during congestion under the segment-wise description are reported.
The concept of area in the fundamental diagram is also introduced,
which facilitates the classification of dynamics during congestion.
We show how the incurred economic loss is related to the macroscopic
variables in the segment-wise description in Sec. \ref{sec:Economic-Loss-Induced}.
Finally, we summarize and discuss the implications of the area enclosed
and various dynamics in congestion.

\section{\label{sec:Description-of-Traffic}Descriptions of Traffic System}

\subsection{Conventional Traffic Models}

The importance of studying highway traffic has long been recognized.
The famous LWR model about traffic on highway was proposed around
1955 by Lighthill, Whitham and Richards \citep{lighthill1955kinematic,richards1956shock}.
It was the most well-known model of traffic. The fundamental assumption
of the model is that flux $\Phi$ and density $\rho$ can be defined
at any point on the highway, and the flux $\Phi$ is a function of
the density $\rho$. The fundamental relation of traffic flow was
defined as
\begin{equation}
\Phi=v\rho,\label{eq:fd}
\end{equation}
where $v$ is interpreted as the average speed of the flow. Numerous
works were developed afterwards \citep{daganzo1995requiem,newell1993simplified,whitham2011linear,nagel2003still}.

There are two phases of traffic systems in the model, the free flow
phase, and the congestion phase according to the density of vehicles,
$\rho$. There is a critical value of density, $\rho_{c}$. When $\rho<\rho_{c}$,
the system is in the free flow phase. The flow speed $v$ is a constant,
corresponding to a free-will speed of vehicles, $v=v_{\text{free}}$.
The flux $\Phi$ increases linearly with $\rho$. For $\rho\ge\rho_{c}$,
the system is in the congestion phase. The interaction between vehicles
becomes significant. The flow speed $v$ is a decreasing function
of $\rho$. The flux $\Phi$ decreases with $\rho$. 

If vehicles are perfectly coordinated, they could maintain at almost
the free-will speed independent of $\rho$. The decrease of $v$ implies
an increase of traveling time, and hence an economic loss to the society.

In the LWR model, these phases are defined in the steady state, in
which the speed is steady and uniform for all vehicles. In reality,
the traffic system is not in the steady state because of fluctuations
of driver behaviors and vehicle influx.

However, there are situations that require a non-local description
of the traffic. One situation arises in studying the impact of congestion
to the society. The economic loss due to congestion is proportional
to the total delay incurred during congestion. In the point-wise description,
the delay can only be approximated from the decrease in speed. However,
as congestion is a phenomenon with finite length, this approximation
requires a frequent sampling of points on the highway to be accurate.
On the contrary, in the segment-wise description, the total delay
comes naturally as it is just the increase in traveling time along
the segment. We shall show that it is possible to compute the economic
loss through macroscopic variables in the segment-wise description.

Another situation arises in clarifying the nature of congestion proposed
by various traffic theories. For example, with observations from real
congested traffic patterns, Kerner proposed the notion of synchronized
flow \citep{kerner1996experimental,kerner1998experimental,kerner1999congested,kerner1999physics,kerner2002synchronized,kerner2014probabilistic}.
In this notion, the conventional congestion phase is composed of two
other phases, ``the synchronized flow'' and ``the wide moving jam''.
The two behaviors mainly differ in the property at the downstream
front. The ``wide moving jam'' traffic phase is defined as a moving
jam with a propagating constant-speed downstream front \citep{kerner2005physics}.
The ``synchronized flow'' phase is defined as a complement of the
``wide moving jam'' traffic phase. The location of the downstream
front in the synchronized flow is fixed but the mean velocity of vehicles
in the downstream front is not maintained during the phase. Although
there are already lots of works modeling this three phase traffic
theory \citep{davis2008driver,gao2007cellular,hoogendoorn2008macroscopic,jia2011cellular},
there were criticisms of its inconsistency in the definition of the
synchronized phase \citep{schonhof2009criticism}. It was suggested
that the synchronization of vehicle speed among different lanes was
merely a transient characteristic of the ``synchronized flow'' at
the beginning of the three-phase traffic theory \citep{kerner1996experimental},
but it was found later by Kerner that this behavior could be observed
in congestion \citep{kerner2005physics}. 

These discussions bring up the need of a segment-wise description
in analyzing congestion. It is noted that the two phases introduced
by Kerner require knowledge of behaviors at the downstream front,
and could not be recognized by examining the density and flux at a
point only. In the conventional LWR model, the fundamental relation,
Eq. \eqref{eq:fd}, is only defined at a point, and hence is just
a point-wise description of the traffic system. While settling these
controversies is not the intention of this article, it is essential
that the ultimate solution requires a segment-wise description. On
the other hand, this article focuses more on the dynamical picture
of congestion and identify two different system behaviors.

\subsection{The 1-point Measure and the 2-point Measure}

There are various ways to describe a traffic system. They are mainly
different in definitions of the major observables: density $\rho$,
flux $\Phi$, and speed $v$.

A point-wise description uses information at a fixed location on the
system \citep{lighthill1955kinematic,kerner1996experimental,orosz2010traffic}.
Consider a detector placed at a fixed point $x$ on the highway, and
a vehicle passing through the detector. The information collected
would be a time interval $\left[t_{1},t_{2}\right]$ when the vehicle
is on the detector. Hence, the vehicle would contribute to the flux
at time $t_{1}$. Its speed can be approximated as
\begin{equation}
v_{i}=\frac{L_{\text{veh}}}{\left|t_{2}-t_{1}\right|},
\end{equation}
where $L_{\text{veh}}$ is the typical length of a vehicle. There
is no information of the density $\rho$ but it can be computed through
the Edie's definition \citep{edie1963discussion}. For an observation
time interval $\left[t,t+\varDelta t\right]$. The flux is 
\begin{equation}
\Phi_{1}\left(x,t\right)=\frac{n\left(x,t\right)}{\varDelta t},
\end{equation}
where $n\left(x,t\right)$ is the number of vehicles passing through
$x$ in $\left[t,t+\varDelta t\right]$. The density follows from
the fundamental relation, Eq. \eqref{eq:fd} with the harmonic mean
of speed \citep{edie1963discussion}
\begin{equation}
\rho_{1}\left(x,t\right)=\frac{\Phi_{1}\left(x,t\right)}{v_{\text{har}}\left(x,t\right)},\label{eq:den1}
\end{equation}
\begin{equation}
v_{\text{har}}=\left(\frac{1}{n\left(x,t\right)}\sum_{i=1}^{n\left(x,t\right)}\frac{1}{v_{i}}\right)^{-1}.
\end{equation}

A large number of studies employed this definition of $\left(\rho_{1},\Phi\right)$
\citep{gupta2015phase,jin2013spontaneous,kerner1996experimental,kerner2014probabilistic,kerner2016failure,nagelocd1994traffic,sugiyama2008traffic}.
However, $v_{\text{har}}$ would be greatly affected by low speed
vehicles, and hence fluctuates rapidly in congestion. The computed
density $\rho_{1}$ might be affected. The formula may work well when
the system is steady and uniform. In the case of large fluctuations
in speed, $\rho_{1}$ may not reflect the true condition of the system.

A segment-wise description uses information of vehicles within a segment.
A commonly accepted segment-wise definition of $\Phi$ and $\rho$
is stated in the Highway Capacity Manual \citep{ni2004direct,washington1985highway}.
Consider a road segment $\left[x-L,x\right]$ on the highway and an
observation time interval $\left[t,t+\varDelta t\right]$. The flux
$\Phi_{2}$ is defined to be the number of vehicles passing the downstream
end $x$ within the time interval, while the density $\rho_{2}$ is
defined as the averaged number of vehicles along the road segment
at time $t$. From the definition, $\Phi_{2}$ is the same as that
in the point-wise description,
\begin{equation}
\Phi_{2}\left(x,t\right)=\frac{n\left(x,t\right)}{\varDelta t}.
\end{equation}
The density $\rho_{2}$ is different,
\begin{equation}
\rho_{2}\left(x,t\right)=\frac{\alpha\left(x,t\right)}{L},\label{eq:den2}
\end{equation}
where $\alpha\left(x,t\right)$ is the number of vehicles in $\left[x-L,x\right]$
at a time $t$. 

It is worth mentioning that $\rho_{2}L=\alpha$ is the total number
of vehicles in the segment. Through describing the traffic system
in this way, we are making an analogy with the queuing theory. The
flux in traffic system corresponds to the service rate in the queuing
theory \citep{cooper1981introduction}. With more vehicles accumulated
on the segment, the time to pass through the segment increases. 

In fact, the quantity $\alpha$ is the accumulation introduced by
Daganzo \citep{daganzo2007urban,geroliminis2007macroscopic}. Accumulation
is the number of vehicles in the traffic system. It is a convenient
quantity as it is additive when two or more segments are combined.
A macroscopic fundamental diagram was successfully reproduced by summing
different links on the traffic network in an area. 

The accumulation $\alpha$ can be obtained by taking a snapshot of
the system and counting the number at any given time. It is difficult
and demanding in real practice. Instead, it can be obtained easily
through monitoring detectors at the upstream and downstream ends. 

For a closed road segment, by conservation of flow, the time evolution
of $\alpha$ follows the equation
\begin{equation}
\alpha\left(t\right)=\alpha\left(t_{0}\right)+\int_{t_{0}}^{t}\left(\Phi\left(x-L,t'\right)-\Phi\left(x,t'\right)\right)dt',\label{eq:acc2}
\end{equation}
with $\Phi\left(x-L,t'\right)$ and $\Phi\left(x,t'\right)$ denoting
the flux detected at the upstream and downstream ends, respectively,
and $\alpha\left(t_{0}\right)$ denoting the number of vehicles in
the system initially. It would be negligible by starting the computation
of $\alpha$ when $\alpha\left(t_{0}\right)$ is close to $0$ by
common sense, such as $3:00$ a.m.

Hence, the definition of $\left(\rho_{2},\Phi\right)$ requires information
from two locations, $\left(x-L\right)$ and $x$. 

To emphasize the difference between the two descriptions, they are
hereafter named as the 1-point measure and the 2-point measure, respectively.
The 1-point measure uses the information of appearance and speed at
the downstream end, while the 2-point measure uses the information
of appearance at the upstream end and the downstream end.

\subsection{Comparison in Simulations}

The two descriptions are equivalent when the system is steady and
uniform. Suppose $N$ vehicles move with speed $v$ and span the system
evenly. Then the average number of vehicles per length is just $\frac{\alpha}{L}$
from the 2-point measure. The number of vehicles in the region $\left[x-\varDelta x,x\right]$
is $\frac{\alpha}{L}\varDelta x$. Then for a time interval $\varDelta t$,
the number of vehicles passing $x$ is the number of vehicles within
the region $\left[x-v\varDelta t,x\right]$, 
\begin{equation}
n=\frac{\alpha}{L}v\varDelta t.
\end{equation}
Hence, from Eq. \eqref{eq:den1}, 
\begin{equation}
\rho_{1}=\frac{n}{v\varDelta t}=\frac{\alpha}{L}=\rho_{2}.
\end{equation}
This steady and uniform condition holds in the free flow phase, but
not in the congestion phase. 

We compare the two descriptions through simulations of the optimal
velocity model (OVM). The purpose of doing these simulations is that
we can control the parameters in simulations and hence fix whether
the system is in the free flow phase or in the congestion phase.

The OVM is well-studied in various literature \citep{orosz2009exciting,orosz2010traffic,sugiyama1999optimal}.
It is a single lane model. Every vehicle has an optimal velocity to
achieve. This optimal velocity $V\left(h\left(t\right)\right)$ depends
on the headway distance $h\left(t\right)$ only and is homogeneous
among vehicles. The headway distance $h$ is defined as the distance
between the center of mass of vehicle with that of the one ahead.
There are two contributions to the acceleration. One is the relaxation
of the current speed $v$ to $V\left(h\right)$. The other one depends
on the velocity of the vehicle ahead as seen by the driver. Drivers
tend to follow the speed of the vehicle ahead.

The OVM can be summarized by the following sets of differential equations,
\begin{equation}
\begin{cases}
\dot{h}_{i}\left(t\right)=v_{i+1}\left(t\right)-v_{i}\left(t\right)\\
\dot{v}_{i}\left(t\right)=\frac{1}{T}\left(V\left(h_{i}\left(t-\tau\right)\right)-v_{i}\left(t-\kappa\right)\right)+\beta\dot{h}_{i}\left(t-\sigma\right)\\
V\left(h\right)=v_{\text{max}}\begin{cases}
0 & if\;h<L_{\text{veh}}\\
\frac{\left(h-L_{\text{veh}}\right)^{3}}{8L_{\text{veh}}^{3}+\left(h-L_{\text{veh}}\right)^{3}} & if\;h\ge L_{\text{veh}}
\end{cases}
\end{cases},
\end{equation}
where $L_{\text{veh}}=7.5\:\text{m}$ is the typical length of vehicles
\citep{nagel1992cellular}, $v_{\text{max}}=110$ $\mathrm{kmh^{-1}}$.
$\tau$, $\kappa$ and $\sigma$ models the reaction times of driver
to the observables $h_{i}$, $v_{i}$, $\dot{h}_{i}$, respectively.
$T$ and $\beta$ are which can be determined by the following. 

Drivers have different reaction times on different quantities in general.
We employ the human driver set-up. Drivers take time to respond to
the change in relative position, but they could react instantaneously
to self motion. The parameters are then set to be $\tau=\sigma=\frac{2L_{\text{veh}}}{v_{max}}=0.4904\:\mathrm{s}$,
$\kappa=0$. $T=\frac{2L_{\text{veh}}}{v_{max}}=0.4904\:\mathrm{s}$,
$\beta=\frac{v_{\text{max}}}{10L_{\text{veh}}}=0.407\:\mathrm{s}^{-1}$
\citep{orosz2010traffic}.

We simulate the model on a system of length $L_{\text{sim}}$ , with
periodic boundary condition. Random sequential update is applied.
Vehicles are distributed uniformly on the segment with random initial
speeds. By varying $L_{\text{sim}}$, the system evolves from the
free flow phase to the congestion phase.

\begin{figure}
\begin{centering}
\includegraphics[scale=0.4]{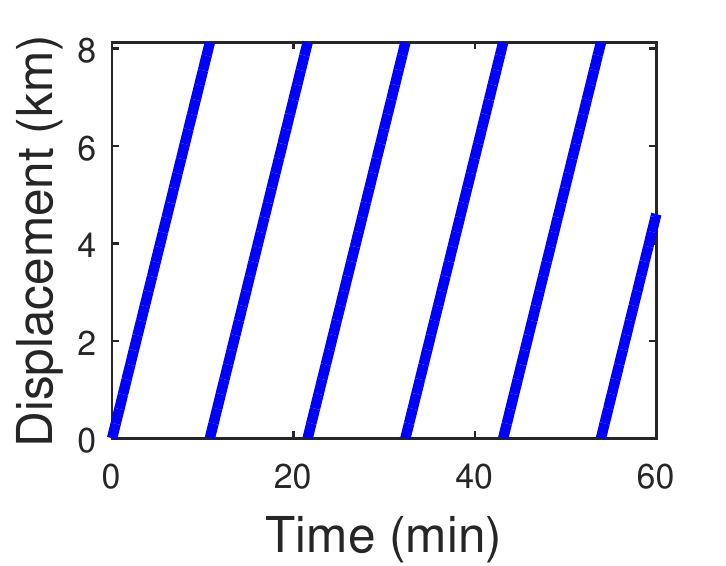}\includegraphics[scale=0.4]{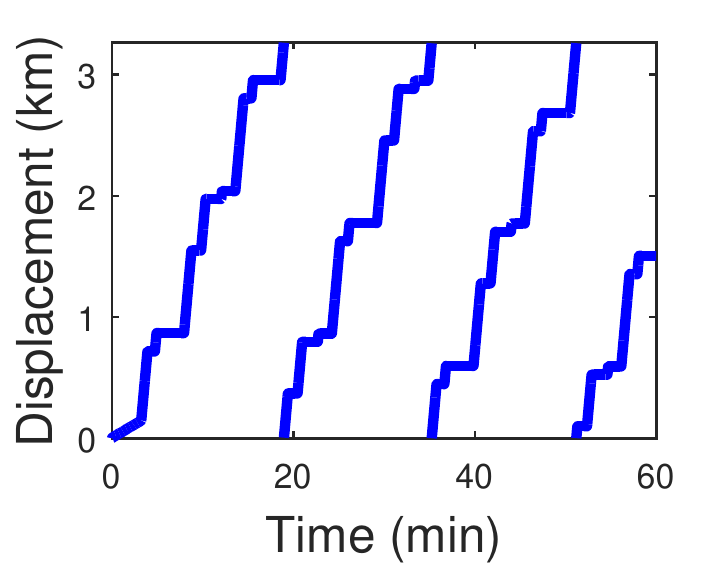}\includegraphics[scale=0.4]{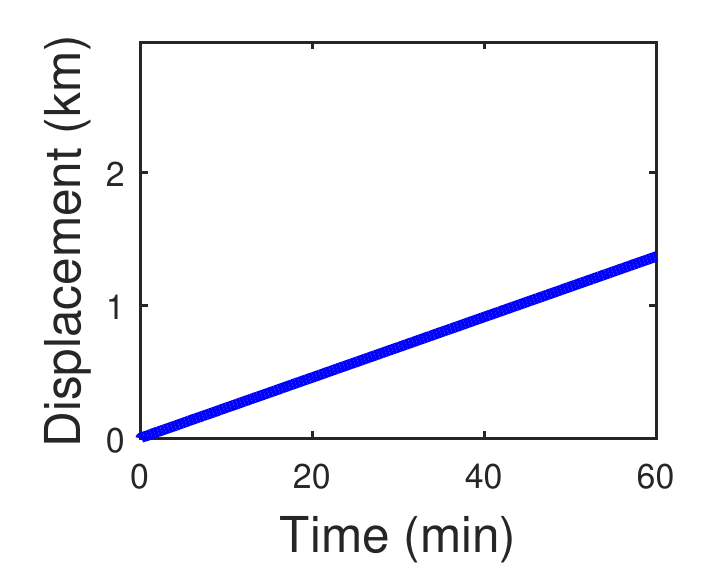}
\par\end{centering}
\caption{\label{fig:OVM}Trajectories of a vehicle from OVM with $N=250$,
in free flow phase $L_{\text{sim}}=8.13\:\mathrm{km}$ (left), intermediate
phase $L_{\text{sim}}=3.26\:\mathrm{km}$ (middle), slowly moving
jam phase $L_{\text{sim}}=2.98\:\mathrm{km}$ (right). The range of
y-axis is the same as $L_{\text{sim}}$.}
\end{figure}
Figure \ref{fig:OVM} shows the trajectories of one vehicle in different
situations. When $L_{\text{sim}}$ is large, the system is in the
free flow phase. Vehicles do not interact with each other. The trajectories
are straight and their speeds are steady. When $L_{\text{sim}}$ decreases
to the intermediate regime ($L_{\text{sim}}=3.05\:\mathrm{km}$),
vehicles are forced to interact. A vehicle stops if it is too close
to the vehicle ahead, and accelerates when there is free space. This
results in winkles on the trajectories. The phenomenon is usually
referred to as a stop-and-go wave in the literature. The system is
in the congestion phase as it is not in the free flow phase anymore.
When $L_{\text{sim}}$ continues to decrease, vehicles become too
close to each other. All vehicles could not accelerate too much and
are synchronized. The trajectory is a straight line without winkles.
Vehicles move together at a steady and extremely low speed.

To compare the 1-point measure and 2-point measure, we consider a
subsystem of the whole simulation environment. Let there be two detectors
collecting information of passing vehicles at $x_{1}$ and $x_{2}$,
with $x_{2}>x_{1}$. Vehicles are considered to be traveling from
$x_{1}$ to $x_{2}$. The 1-point measure uses the appearance and
speed of vehicles from the detector at $x_{2}$ only, while the 2-point
measure uses the appearance of vehicles at $x_{1}$ and $x_{2}$.
The flux going out of the subsystem $\left[x_{1},x_{2}\right]$ is
the rate of vehicles detected at $x_{2}$.

\begin{figure}
\begin{centering}
\includegraphics[scale=0.7]{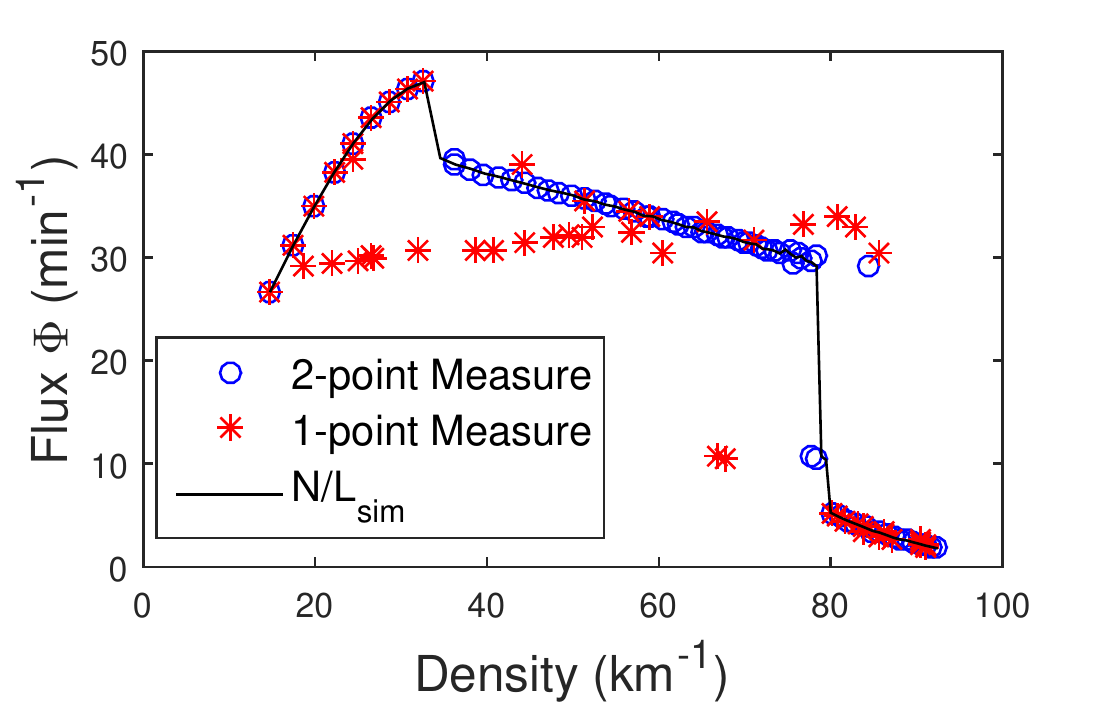}
\par\end{centering}
\caption{\label{fig:OVMfd}Fundamental diagram from 1-point measure (red star)
and 2-point measure (blue circle) of the optimal velocity model by
varying segment length $L_{\text{sim}}\gg1\;\mathrm{km}$, with $\varDelta t=1\;\mathrm{min},$
$\left\Vert x_{2}-x_{1}\right\Vert =0.08L_{\text{sim}}$. Each point
is an average over the time series after transient. $\left\{ \frac{N}{L_{\text{sim}}},\Phi\right\} $
is plotted in black line as a reference. Vehicles are initially distributed
evenly on the road segment, with normalized inital speed $v$ following
a normal distribution, $\frac{v}{v_{\text{max}}}\sim N\left(\frac{1}{2},\frac{1}{6}\right)$.}
\end{figure}
 Figure \ref{fig:OVMfd} shows the fundamental diagram generated from
major observables of the two descriptions. We also plot $\left\{ \frac{N}{L_{\text{sim}}},\Phi\left(x_{2}\right)\right\} $
on the fundamental diagram. Since $N$ and $L_{\text{sim}}$ are parameters
that control the behaviors of the model, $\frac{N}{L_{\text{sim}}}$
could be treated as a reference density of the system. The two descriptions
coincide with the reference value when $L_{\text{sim}}$ is either
too large or too small. The 1-point measure deviates from the reference
line for intermediate $L_{\text{sim}}$. This is expected since the
steady and uniform conditions do not hold for intermediate $L_{\text{sim}}$.
The computation of density $\rho_{1}$ depends on the speed of vehicles
at $x_{2}$, from Eq. \eqref{eq:den1}. As shown in the trajectory
in Fig. \ref{fig:OVM}, the speed fluctuates in between two extremes,
fast or stationary. When the winkles are upstream of the detector,
only fast vehicles passing the detector contribute to the computation
of $\rho_{1}$. The density would be underestimated. On the other
hand, when the winkles are on the detector, only slow vehicles contribute.
It results in an overestimation of the density. Depending on the time
at which a winkle is on the detector, $\rho_{1}$ would deviate from
the reference value in different ways. This shows that the 1-point
measure is not robust to congestion behaviors. In contrast, $\rho_{2}$
is robust to this fluctuation of vehicle speed, and coincides with
the reference line. It is noted that the result is independent of
the length of the subsystem, unless it is too small.

The purpose of a fundamental diagram is to determine the phase of
a traffic system. A system is said to be congested when $\rho>\rho_{c}$.
Due to speed fluctuations, for a system in the intermediate regime,
the local density $\rho_{1}$ estimated from the 1-point measure could
be smaller than the critical value, $\rho_{1}<\rho_{c}$. It fails
to indicate what phase the system is currently at. 

From this simple simulation of the optimal velocity model, we show
that there is a difference in point-wise description and segment-wise
description, after the breakdown of the free flow phase. The 2-point
measure is robust to speed fluctuations in congestion and matches
the reference density. This shows that in studying congestion, it
would be more reliable by employing the segment-wise 2-point measure.

\section{\label{sec:Congestion-in-Reality}Congestion in Reality}

We analyze real data from the Taiwan highway system, which consists
of $10$ national highways and numerous provincial highways. There
are $2$ major highways connecting the northern and southern parts
of Taiwan, with a length of about $400\:\text{km}$, respectively.
In $2006$, the Electronic Toll Collection system replaced the traditional
toll stations. Overhead detectors were installed on major highways.
Whenever a vehicle passes under these detectors, its identity, location
and time of appearance would be recorded. This provides a precious
opportunity in studying real traffic systems.

The analysis is performed by applying the 2-point measure. Since velocity
data is not available from the dataset, the 1-point measure is not
applicable. Nevertheless, as demonstrated in the OVM simulation in
the next section, the characterization of congestion using the loop
area enclosed in the fundamental diagram is much more obscured if
1-point measurements are adopted. The density of a segment in between
any two consecutive detectors is computed by Eq. \eqref{eq:acc2}.
There are entrances from the local streets and exits to the local
streets in the segments. We approximate the locations of highway entrance
and exit in a segment to be the locations of the downstream and upstream
ends, respectively. It corresponds to the location of the first and
last detections of the entering or exiting vehicles in the segment,
respectively. As the number of vehicles entering or exiting the segment
is small compared with the number of vehicles traveling across the
segment, their contributions are negligible, and the approximation
is reasonable.

\begin{figure}
\begin{centering}
\includegraphics[scale=0.5]{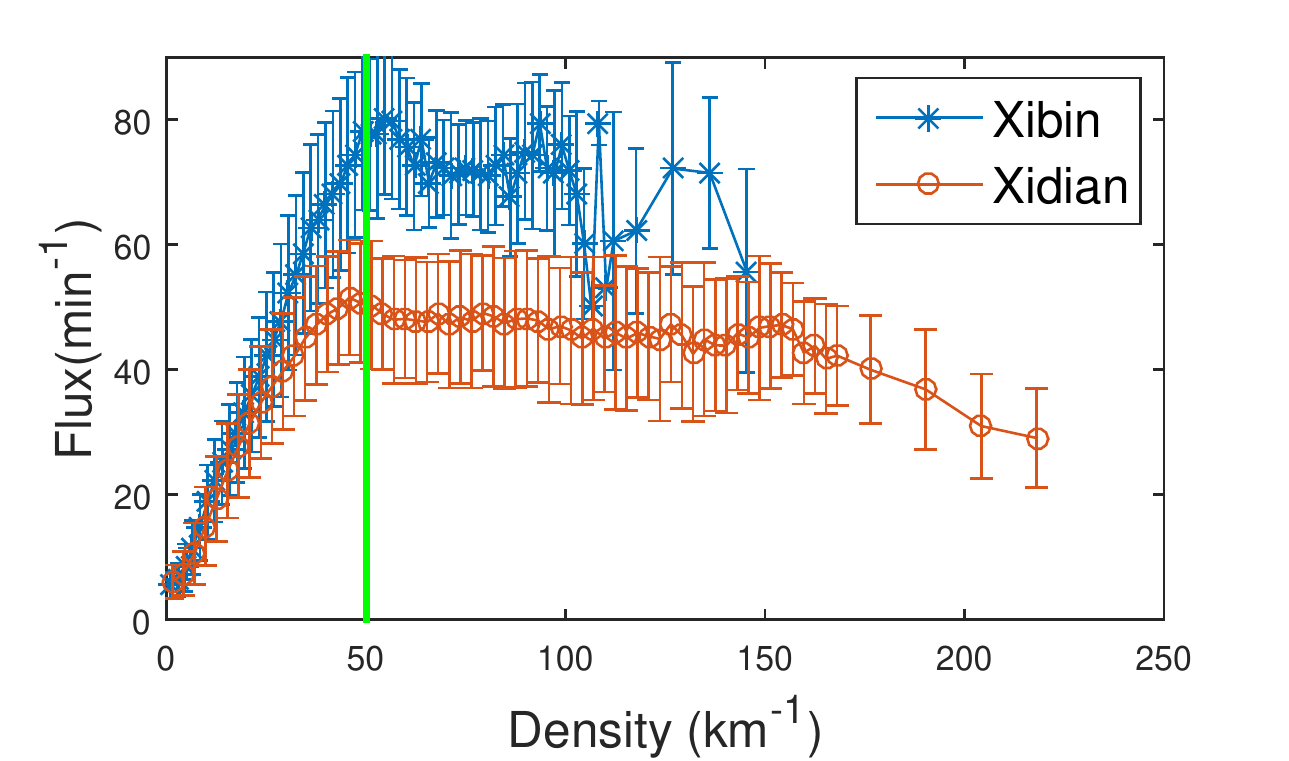}
\par\end{centering}
\caption{\label{fig:Fluxcmp}Flux $\Phi$ versus density $\rho_{2}=\frac{\alpha}{L}$
from Taiwan highway data, collected in February 2016, averaged over
29 days. Data collected in segment from Xibin to Zhunan (blue star)
and from Xindian to Ankeng (red circle). The length $L$ of the segments
are $5.4$ km and $3.6$ km, respectively. The vertical green line
indicates the line $\rho_{2}=50\:\mathrm{km}^{-1}$.}
\end{figure}
 Typically, a truncated fundamental diagram is observed through the
2-point measure for almost all the $300$ segments in the Taiwan highway
system. An example is shown in Fig. \ref{fig:Fluxcmp}. When the density
is small, the system is in the free flow phase. The flux rises linearly
with the growth of density. The free flow phase breaks down for sufficiently
large density as expected, typically at $\rho_{2}\approx50\:\text{km}^{-1}$.
For the Taiwan highway system with $3$ lanes typically, it is equivalent
to a headway distance of about $60$ $\text{m}$. It is roughly the
displacement of a vehicle traveling at the speed limit of $110$ $\text{kmh}^{-1}$
for $2$ $\text{s}$. This is consistent with an experimental finding
by McGehee in 2000 \citep{mcgehee2000driver}, which shows that the
driving reaction time in crash avoidance is about $2.3$ s on average.

The time-averaged congestion behavior agrees with the prediction of
conventional theory. However, this assumes that the variation of flux
at a given density is caused by random fluctuations only. It also
neglects the time correlation in the data. To get better insights
of how congestion unfolds, we consider the time dependence of the
system state in the fundamental diagram.
\begin{figure}
\begin{centering}
\includegraphics[scale=0.4]{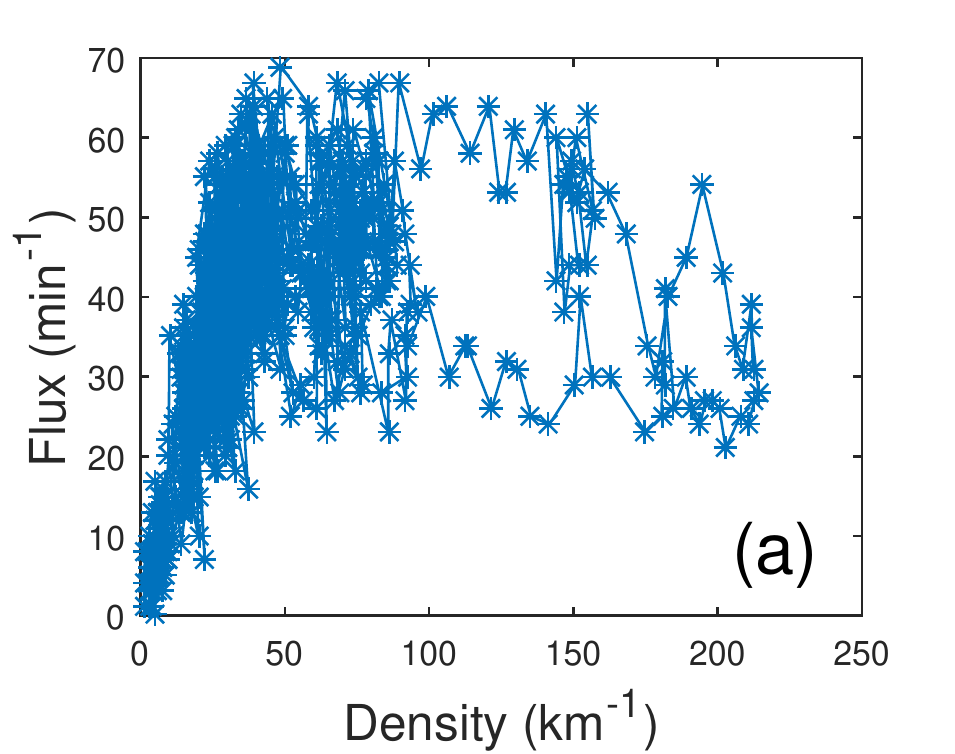}\includegraphics[scale=0.4]{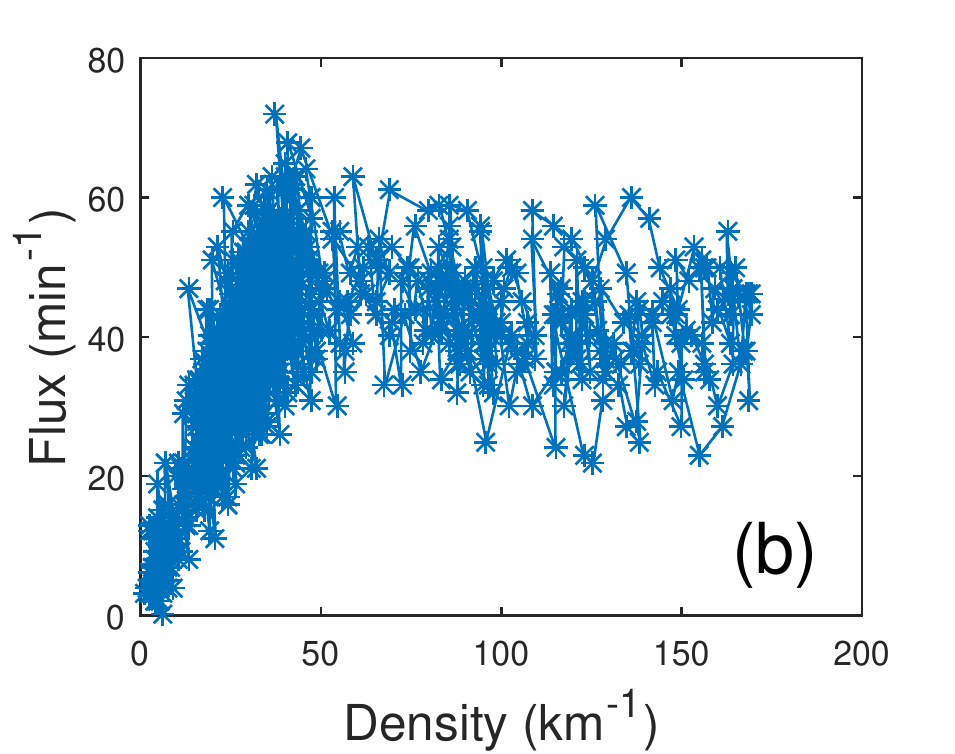}
\par\end{centering}
\caption{\label{fig:The-daily-dynamics}The daily dynamics of the system on
the fundamental diagram. Data collected in the segment from Xindian
to Ankeng, with a length of $3.6$ km (a) on 7-Feb, 2016, and (b)
on 24-Feb, 2016. Each data point is measured in one minute interval,
from $5:00$ a.m. to $11:00$ p.m.}
\end{figure}
 Figure \ref{fig:The-daily-dynamics} shows examples of daily evolution
of the traffic system on the fundamental diagram. It can be observed
that there are two types of evolution after the breakdown of the free
flow phase. A loopy evolution can be observed in Fig. \ref{fig:The-daily-dynamics}(a)
while the flux in Fig. \ref{fig:The-daily-dynamics}(b) remains at
a high level around $40-50$ $\text{min}^{-1}$. 

\begin{figure}
\begin{centering}
\includegraphics[scale=0.5]{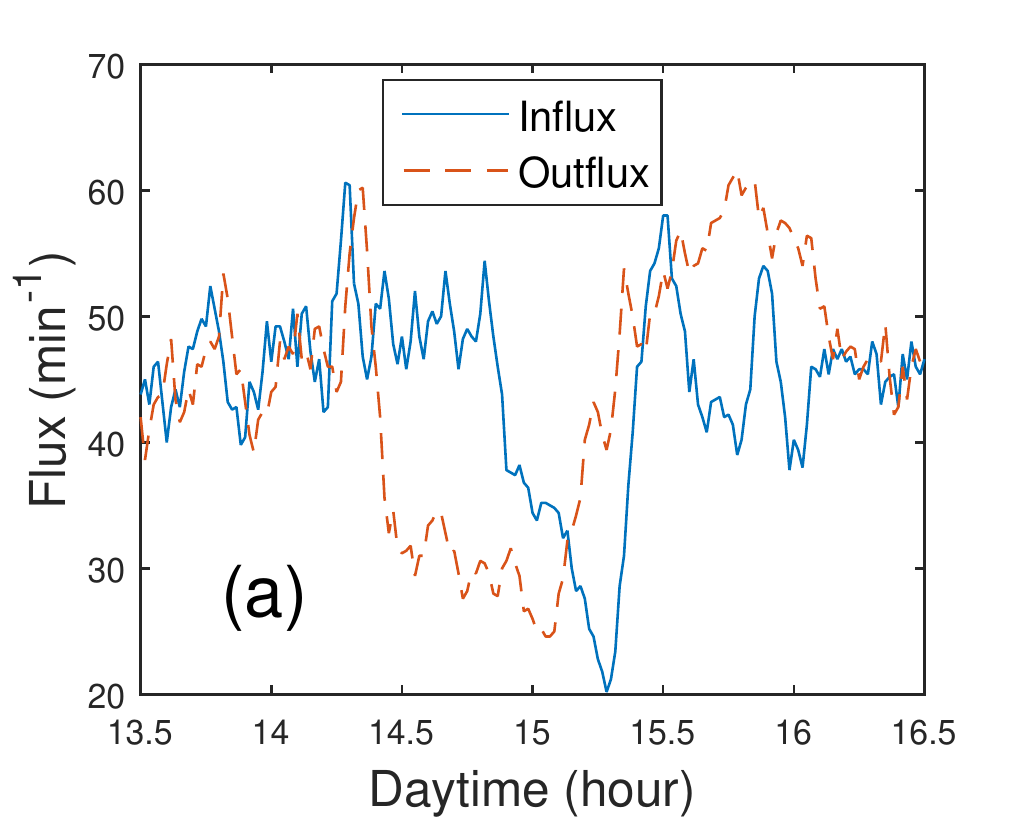}
\par\end{centering}
\begin{centering}
\includegraphics[scale=0.5]{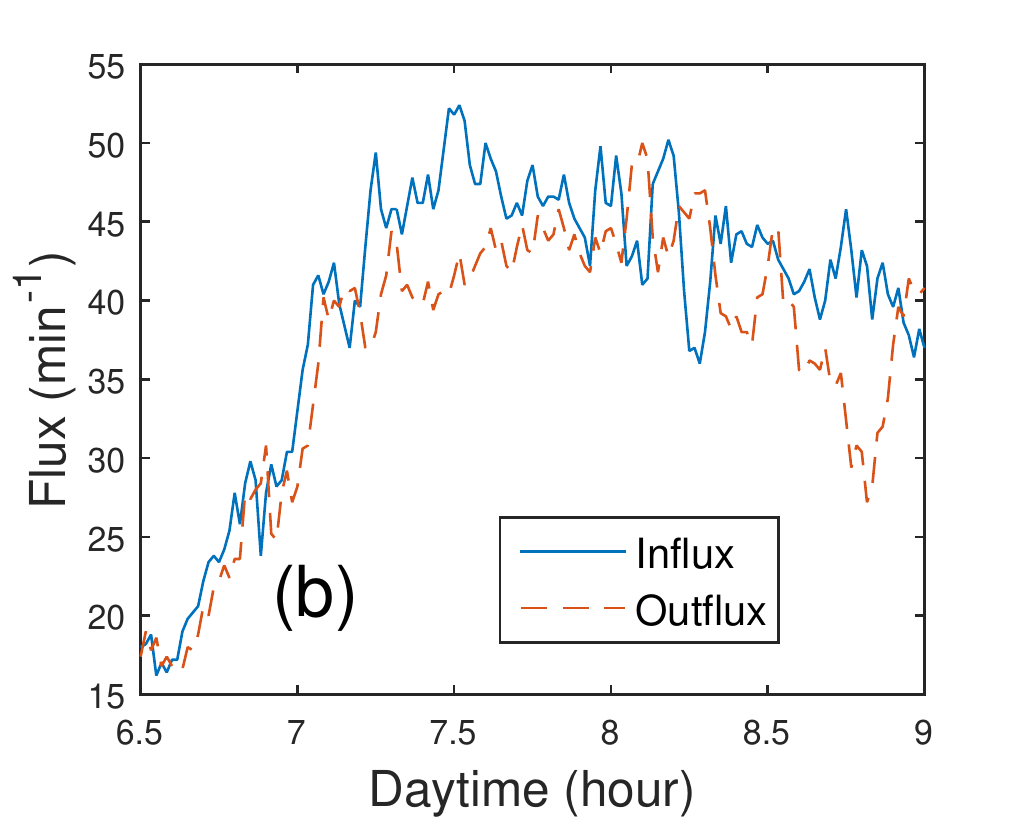}
\par\end{centering}
\caption{\label{fig:inoutflux}The time evolution of influx and outflux of
the segment from Xindian to Ankeng, with a length of $3.6$ km (a)
on 7-Feb, 2016, and (b) on 24-Feb, 2016. }

\end{figure}
 To understand the physical picture of these dynamics, the time series
of the flux at the upstream end (influx) and the downstream end (outflux)
are investigated. The influx always leads the outflux in the free
flow phase due to the finite time for vehicles to traverse the segment.
However, this time correlation may not be maintained during congestion. 

Figure \ref{fig:inoutflux} shows the flux evolution during congestion.
For the loopy dynamics in Fig. \ref{fig:The-daily-dynamics}(a), there
was a sudden drop in outflux in Fig. \ref{fig:inoutflux}(a) around
$02:15$ p.m. The outflux remained low while the number of vehicles
inside the system built up. A significant drop of influx followed
at around $03:00$ p.m. The system recovered to the normal flux level
when the density reached the maximum value of about $200$ vehicles
per km. It can be observed that the influx was sightly leading the
outflux before $02:15$ p.m. Afterwards, this relation was reversed.
The outflux led the influx, indicating a possibility of back propagation
of congestion.

This reverse of time correlation is not observed for the dynamics
in Fig. \ref{fig:The-daily-dynamics}(b). The correlation between
influx and outflux was lost during congestion, starting from around
$07:15$ a.m. This may be explained by stochastic interactions among
individuals. The interaction was internal among the vehicles and hence
did not affect the influx. 

This suggests that there are two possible dynamics after the breakdown
of the free flow phase. To verify whether the two cases are qualitatively
different, or merely correspond to loops of different sizes, we need
to quantify these congestion behaviors. It is noted that they are
common in having an increase in accumulation, and are different in
whether there is a sharp drop in outflux. The product of these two
factors approximates the area enclosed by the dynamics during congestion
in the fundamental diagram. Therefore, it is natural to use the area
enclosed to quantify these behaviors.

Suppose the system experiences a loopy dynamics during a time interval
$\left[t_{a},t_{e}\right]$, the area enclosed by the trajectory can
be computed by
\begin{equation}
A\left(t_{a},t_{e}\right)=\oint\Phi_{\text{out}}\left(t\right)d\rho_{2}\left(t\right),\label{eq:areaenc}
\end{equation}
where the closed path integral is done along the trajectory of the
dynamics in the fundamental diagram, with the two end points being
connected. With assumptions on the behaviors at the end points $t=t_{a},t_{e}$,
and some approximations, it can be shown (in Appendix \ref{sec:areacorrl})
that 
\begin{eqnarray}
A\left(t_{a},t_{e}\right)\approx &  & \frac{\left|t_{e}-t_{a}\right|}{L}\label{eq:areaenccorr}\\
 &  & \times\left(\text{cov}_{\left[t_{a},t_{e}\right]}\left(\Phi_{\text{in}},\Phi_{\text{out}}\right)-\text{cov}_{\left[t_{a},t_{e}\right]}\left(\Phi_{\text{out}},\Phi_{\text{out}}\right)\right),\nonumber 
\end{eqnarray}
where 
\begin{eqnarray}
\text{cov}_{\left[t_{a},t_{e}\right]}\left(A,B\right)= &  & \frac{1}{\left|t_{e}-t_{a}\right|}\\
 &  & \times\int_{t_{a}}^{t_{e}}\left(A\left(t\right)-\bar{A}\left(t\right)\right)\left(B\left(t\right)-\bar{B}\left(t\right)\right)dt,\nonumber 
\end{eqnarray}
 denotes the covariance of the two time series $A\left(t\right)$
and $B\left(t\right)$ in the time interval $\left[t_{a},t_{e}\right]$.

As $\text{cov}_{\left[t_{a},t_{e}\right]}\left(\Phi_{\text{out}},\Phi_{\text{out}}\right)$
is always positive, and an approximation of $\overline{\Phi}_{\text{in}}\approx\overline{\Phi}_{\text{out}}$,
\begin{equation}
\text{cov}_{\left[t_{a},t_{e}\right]}\left(\Phi_{\text{in}},\Phi_{\text{out}}\right)\le\text{cov}_{\left[t_{a},t_{e}\right]}\left(\Phi_{\text{out}},\Phi_{\text{out}}\right).
\end{equation}
Hence, $A\left(t_{a},t_{e}\right)<0$. The sign of the area enclosed
$A\left(t_{a},t_{e}\right)$ indicates the orientation of the loop.
This matches the physical picture as described in Fig. \ref{fig:The-daily-dynamics}(a).
The loops are typically anticlockwise, starting with a sudden drop
of flux and then an increase of accumulation.

\begin{figure}
\begin{centering}
\includegraphics[scale=0.7]{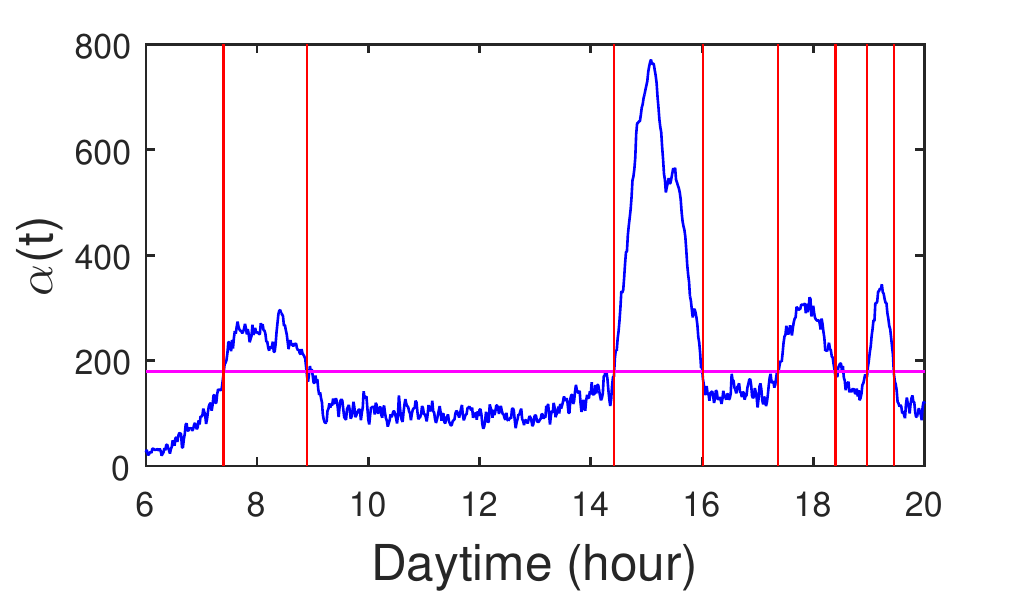}
\par\end{centering}
\caption{\label{fig:Illustration-of-congested}Illustration of congested time
intervals $\left[t_{a},t_{e}\right]$ (indicated by the pairs of red
lines), selected by the criteria: (1) $\forall t\in\left(t_{a},t_{e}\right),\alpha\left(t\right)>\alpha_{c}$;
(2) $\alpha\left(t_{a}\right),\alpha\left(t_{e}\right)\le\alpha_{c}$
. The purple line indicates the threshold, $y=\alpha_{c}$. Data collected
from the system, Xindian to Ankeng, on 16-Feb, 2016 with a length
$L=3.6\text{ km}$.}
\end{figure}

To investigate whether there are really two types of dynamics, we
analyze data from a large number of segments in Taiwan. To employ
Eq. \eqref{eq:areaenc}, the time interval $\left[t_{a},t_{e}\right]$
needs to be defined. As we are investigating the dynamics after the
breakdown of free flows, there is a natural threshold $\alpha_{c}=L\times50\:\text{km}^{-1}$as
the breakdown accumulation. For a time interval $\left[t_{a},t_{e}\right]$
to be identified as the congestion time interval, it must satisfy
the following criteria: (1) $\forall t\in\left(t_{a},t_{e}\right),\alpha\left(t\right)>\alpha_{c}$;
(2) $\alpha\left(t_{a}\right),\alpha\left(t_{e}\right)\le\alpha_{c}$.
An example of the time intervals identified by the criteria is shown
in Fig. \ref{fig:Illustration-of-congested}.

In addition to the area enclosed in $\left[t_{a},t_{e}\right]$, the
range of drop of the outflux in $\left[t_{a},t_{e}\right]$, $\varDelta\Phi\left(t_{a},t_{e}\right)$
is also recorded. It is defined as
\begin{equation}
\varDelta\Phi\left(t_{a},t_{e}\right)=\max_{t\in\left[t_{a},t_{e}\right]}\left(\Phi_{\text{out}}\left(t\right)\right)-\min_{t\in\left[t_{a},t_{e}\right]}\left(\Phi_{\text{out}}\left(t\right)\right).
\end{equation}

\begin{figure}
\begin{centering}
\includegraphics[scale=0.5]{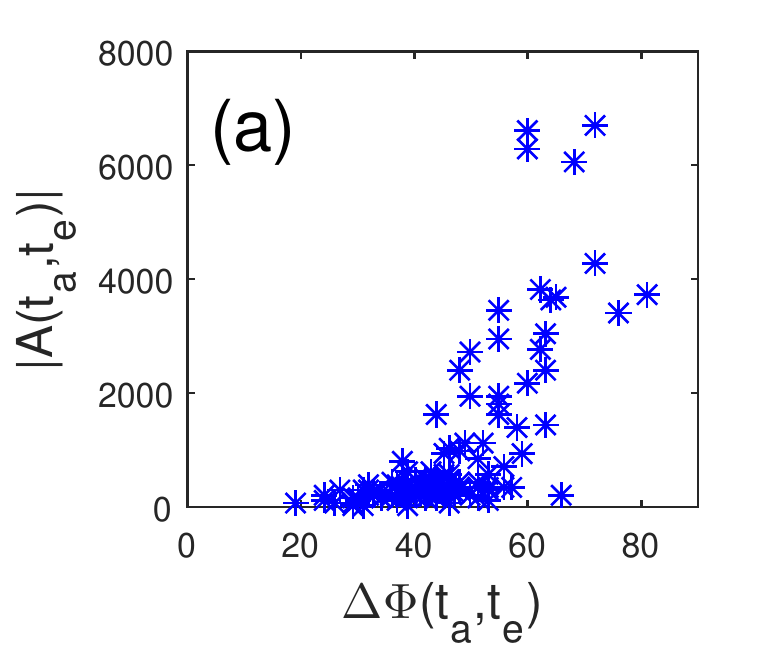}\includegraphics[scale=0.5]{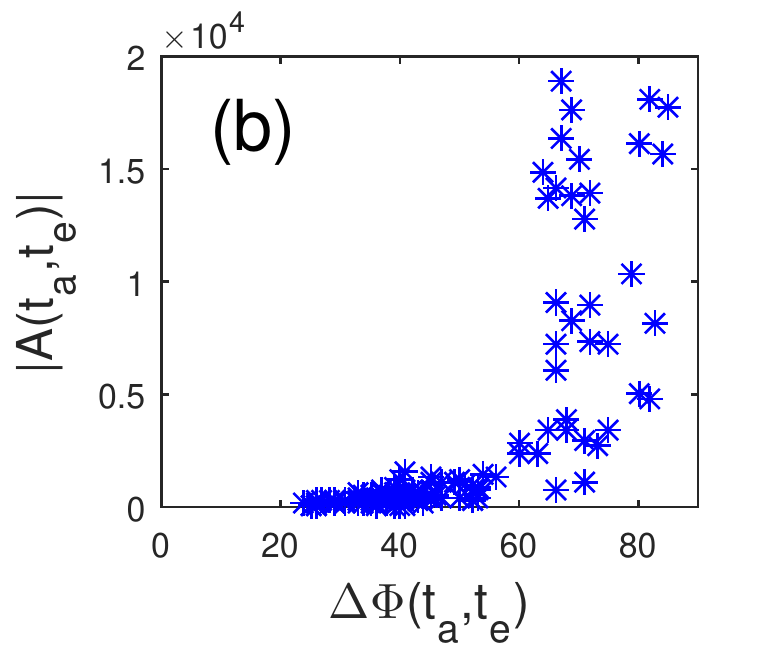}
\par\end{centering}
\caption{\label{fig:Transition}The area enclosed in the congestion $A\left(t_{a},t_{e}\right)$
versus the drop of the outflux in the congestion $\varDelta\Phi\left(t_{a},t_{e}\right)$.
Data collected from the system in Feb, 2016, (a) from Wugu to Sanchong
with a length $L=3.7\text{ }\mathrm{km}$; (b) from Neili to Jongli
with a length $L=2.5$ km.}
\end{figure}
 Figure \ref{fig:Transition} shows how the area enclosed $A\left(t_{a},t_{e}\right)$
depends on the range of outflux drop in different segments. When the
drop is small, the area enclosed is also small. We may classify this
dynamics as the one illustrated in Fig. \ref{fig:The-daily-dynamics}(b).
When the drop of the outflux exceeds a threshold, $A\left(t_{a},t_{e}\right)$
increases sharply. This indicates the occurrence of loopy dynamics.
The threshold for the onset of loopy dynamics varies for different
segments, which may depend on individual characteristics. The drastic
increase of $A\left(t_{a},t_{e}\right)$ with respect to $\varDelta\Phi\left(t_{a},t_{e}\right)$
supports the existence of two dynamics after the breakdown of the
free flow phase. We classify the dynamics in Figs. \ref{fig:The-daily-dynamics}(a)
and (b) as ``serious congestion'' and ``moderate flow'', respectively. 

\section{\label{sec:Economic-Loss-Induced}Economic Loss Incurred in Congestion}

\subsection{Economic Loss in Reality}

Whenever there is congestion, there is an increase of traveling time
across the system. From drivers' point of view, there is always a
better way to spend this delay than suffering from congestion. From
society's point of view, wasting time on transportation does not produce
any value to the economy of the society. Hence, the total incurred
delay can be used to represent the economic loss during congestion.

The latency is defined as the traveling time across the system. Denote
the minimum latency across the system as $l_{\text{free}}$. When
a driver suffers from congestion, the latency increases from $l_{\text{free}}$
to $l$. The monetary loss incurred on the society is $\beta\left(l-l_{\text{free}}\right)$,
where $\beta$ is the value of time. It might vary among individuals,
but for simplicity, it is set to be homogeneous and $\beta=1$. The
economic loss of the society $E_{\text{loss}}$ is 
\begin{equation}
E_{\text{loss}}=\sum_{i=1}^{N_{\text{tot}}}\beta_{i}\left(l_{i}-l_{\text{free}}\right)=\sum_{i=1}^{N_{\text{tot}}}\left(l_{i}-l_{\text{free}}\right),
\end{equation}
where $l_{i}$ is the latency of the $i$th driver, and $N_{\text{tot}}$
is the number of vehicles being affected by congestion. $N_{\text{tot}}$
can be computed by consideration of the influx in the congestion time
interval,
\begin{equation}
N_{\text{tot}}=\int_{t_{a}}^{t_{e}}\Phi_{\text{in}}\left(t\right)dt.
\end{equation}
Let $l\left(t\right)$ be the average latency of vehicles entering
a segment at time $t$. The economic loss incurred by congestion in
$\left[t_{a},t_{e}\right]$ can be expressed in terms of $\Phi_{\text{in}}$,
\begin{equation}
E_{\text{loss}}=\int_{t_{a}}^{t_{e}}\left(l\left(t\right)-l_{\text{free}}\right)\Phi_{\text{in}}\left(t\right)dt.
\end{equation}

To compute $l_{i}$ or $l\left(t\right)$, the information of the
entrance time and exit time of the $i$th individual is required.
A tracking of vehicle's identity across different detectors is needed.
The conventional detectors on the highway only measure the number
of vehicles passing through, without tracing the identity of vehicles.
In contrast, the electronic toll collection system enables us to track
the latency of individual vehicles.

\begin{figure}
\begin{centering}
\includegraphics[scale=0.5]{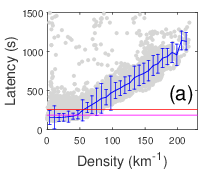}\includegraphics[scale=0.5]{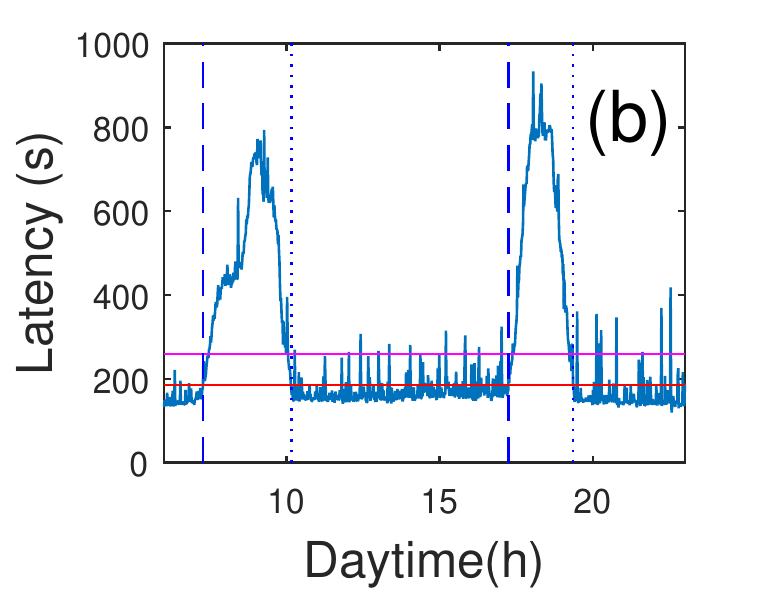}
\par\end{centering}
\caption{\label{fig:l_t_e.g.}(a) Latency $l\left(t\right)$ versus the scaled
accumulation $\frac{\alpha\left(t\right)}{L}$ on the same segment.
The horizontal lines indicate the expected latency if a vehicle travel
at $50\text{ km/h}$ (purple), and $70\text{ km/h}$ (red). (b) An
example of the time evolution of $l\left(t\right)$. Data collected
from the system on 24-Feb, 2016, from Xindian to Ankeng with a length
$L=3.6\text{ km}$. The vertical lines indicate the congestion time
interval $\left[t_{a},t_{e}\right]$ identified by the conditions.
The dashed lines and the dotted lines mark the start and end of the
time intervals, respectively. }
\end{figure}
 The congestion time interval $\left[t_{a},t_{e}\right]$ could also
be defined using $l\left(t\right)$. The system is defined to be congested
in a time interval $\left[t_{a},t_{e}\right]$ if the following conditions
hold. (1) $\left\Vert t_{e}-t_{a}\right\Vert >T$, for $T>0$; (2)
$\forall t\in\left(t_{a},t_{e}\right),\;l\left(t\right)>l_{\text{ref}}$;
(3) $l\left(t_{a}\right),l\left(t_{e}\right)<l_{\text{ref}}$; (4)
$\exists t\in\left(t_{a},t_{e}\right),\;l\left(t\right)>l_{\text{jam}}$.
Condition (2) and (3) are equivalent to the conditions listed in Sec.
\ref{sec:Congestion-in-Reality}. Conditions (1) and (4) are added
in consideration of the rapid fluctuations of the latency time series.

The parameters in these conditions are set in the following way. The
parameter $l_{\text{ref}}$ is the reference latency, determined by
the average latency before breakdown as illustrated in Fig. \ref{fig:l_t_e.g.}(a).
The parameter $l_{\text{jam}}$ is set in order to filter out points
in which the latency is large due to fluctuations, 
\begin{equation}
l_{\text{jam}}=l_{\text{ref}}+\sigma_{l_{\text{ref}}},
\end{equation}
with $\sigma_{l_{\text{ref}}}$ being the standard deviation of $l\left(t\right)$
at the breakdown point. The time interval $T$ is set to a sufficiently
large value to identify a sustained congestion period despite rapid
fluctuations of the latency $l\left(t\right)$.

Figure \ref{fig:l_t_e.g.}(b) shows an example of the identificaion
of the congestion time interval through $l\left(t\right)$. In this
example, the values of $l_{\text{ref}}$ and $l_{\text{jam}}$ correspond
to the expected traveling time for a vehicle of speed $70$ $\text{km/h}$
and $50$ $\text{km/h}$, respectively in Fig. \ref{fig:l_t_e.g.}(a),
and $T$ is set to be $10$ min.

To compute $E_{\text{loss}}$, we note that the time saved by traveling
beyond the speed limit should not be counted as a benefit to the society.
Hence the time saved for traveling time shorter than $l_{\text{ref}}$
is discarded in the following empirical formula for $E_{\text{loss}}$,
\begin{equation}
E_{\text{loss}}=\int_{t_{a}}^{t_{e}}\text{max}\left(l\left(t\right)-l_{\text{ref}},0\right)\Phi_{\text{in}}\left(t\right)dt.
\end{equation}

The evolution of the loopy dynamics consists of the following stages.
There is an initial drop of the outflux, followed by an increase in
the density. When the system recovers, the outflux increases. followed
by a decrease in the density back to the normal state. In reality,
the transitions between these stages is not sharp, but by assuming
sharp transitions the area becomes rectangular. Hence we are able
to relate the economic loss $E_{\text{loss}}$ and the area enclosed
$A\left(t_{a},t_{e}\right)$ (shown in Appendix \ref{sec:Approximation-of-the}),
\begin{equation}
\frac{E_{\text{loss}}}{N_{\text{tot}}}\approx\frac{L}{2f\Phi_{\text{acc}}^{2}}A\left(t_{a},t_{e}\right),\label{eq:Eloss}
\end{equation}
where the economic loss is scaled by $N_{\text{tot}}$, the total
number of vehicles using the segment during the congestion period,
and $f$ is the fractional decrease of the flux during the congestion
time interval. This fractional decrease reflects the decrease of highway
capacity as part of the highway may be blocked. $\Phi_{\text{acc}}$
indicates the influx to the system when the loopy dynamics starts.
Hereafter the area $A\left(t_{a},t_{e}\right)$ will be referred to
as the loss area. 

\begin{figure}
\begin{centering}
\includegraphics[scale=0.5]{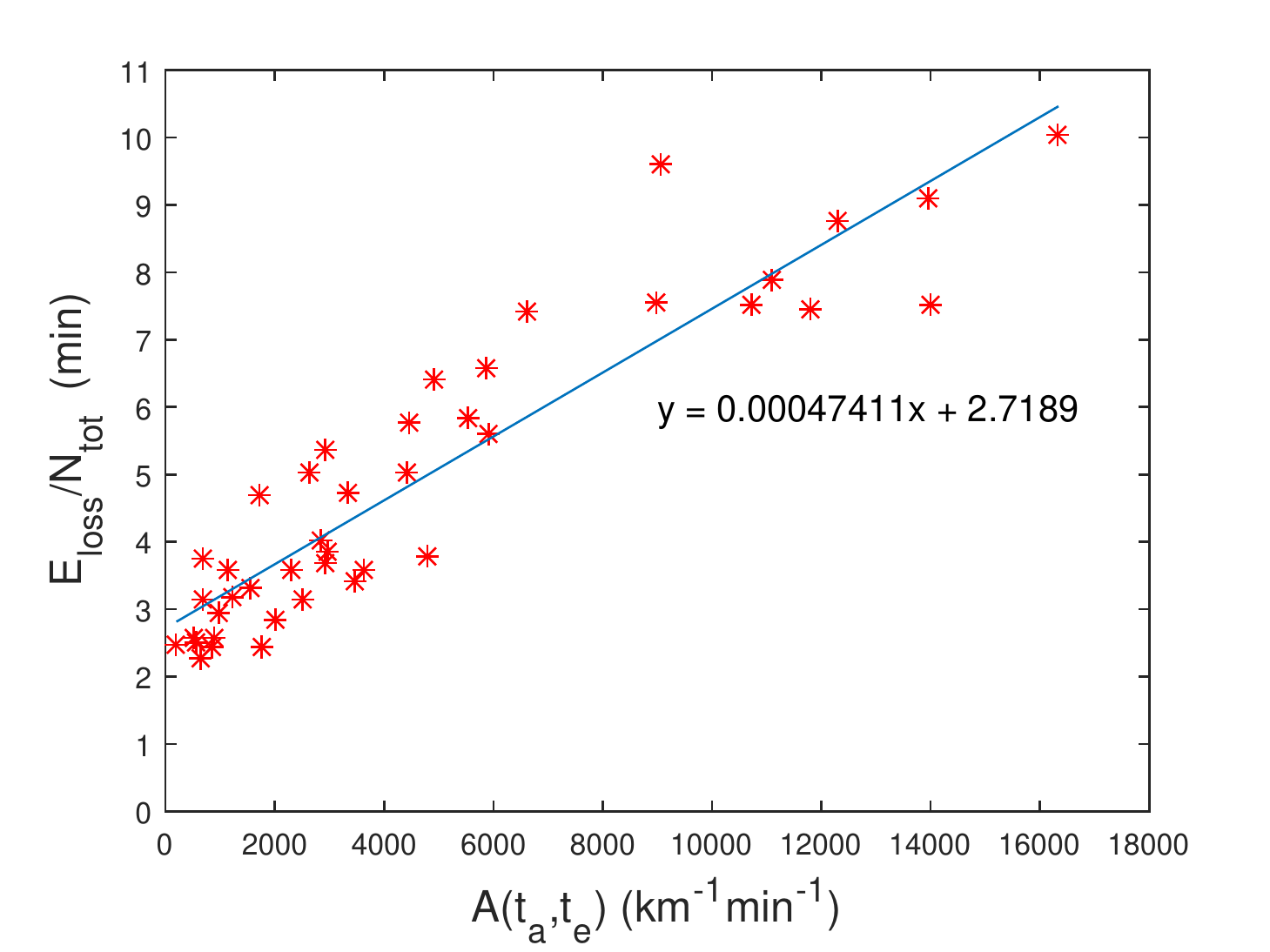}
\par\end{centering}
\caption{\label{fig:Elossvsarea}Economic loss per vehicles $E_{\text{loss}}/N_{\text{tot}}$
versus the area enclosed $A\left(t_{a},t_{e}\right)$. Data collected
on the segment from Xindian to Ankeng with a length $L=3.6\text{ km}$.
The blue solid line is a linear fitting of the data, with slope $m=4.47\times10^{-4}\text{ km}$-$\text{min}^{2}$
and intercept $c=2.72\text{ min}$. }
\end{figure}
 Figure \ref{fig:Elossvsarea} shows how the economic loss per vehicles
$\frac{E_{\text{loss}}}{N_{\text{tot}}}$ is related with the area
enclosed $A\left(t_{a},t_{e}\right)$. A clear linear relation can
be observed. By a linear fitting, 
\begin{equation}
\frac{E_{\text{loss}}}{N_{\text{tot}}}=mA\left(t_{a},t_{e}\right)+c,
\end{equation}
with $m=4.47\times10^{-4}\text{ km}$-$\text{min}^{2}$ and $c=2.72\text{ min}$.
$c\ne0$ because Eq. \eqref{eq:Eloss} apply for loopy dynamics only.
From the criteria listed, $E_{\text{loss}}$ must be non-zero. The
intercept may correspond to the economic loss due to moderate flows
in Fig. 4(b), since their loss areas vanish. We can use the slope
$m$ to approximate $\Phi_{\text{acc}}$ averaged among different
loopy dynamics by Eq. \eqref{eq:Eloss}, $\left\langle \Phi_{\text{acc}}\right\rangle =\sqrt{\frac{L}{2fm}}=72.7\:\text{min}^{-1}$
where $f\approx0.7$ from real data, approximated by 
\begin{equation}
f=\frac{\underset{t\in\left[t_{a},t_{e}\right]}{\max}\left(\Phi_{\text{in}}\left(t\right)\right)-\underset{t\in\left[t_{a},t_{e}\right]}{\text{min}}\left(\Phi_{\text{out}}\left(t\right)\right)}{\underset{t\in\left[t_{a},t_{e}\right]}{\max}\left(\Phi_{\text{in}}\left(t\right)\right)}.
\end{equation}
$\Phi_{\text{acc}}$ has the same order of magnitude as the typical
flux before breakdown of this segment, which is about $50\:\text{min}^{-1}$
shown in Fig. \ref{fig:Fluxcmp}. The deviation may arise from the
approximation in the derivation of Eq. \eqref{eq:Eloss}, where we
have neglected the interactions between vehicles. For example, during
an accident, vehicles traveling on the lane with obstacles ahead would
try to switch to other lanes. This interaction between vehicles from
different lanes would cause further delays. 

Hence, the economic loss per vehicle is highly correlated with the
area enclosed. We are able to determine the average delay suffered
by the drivers during congestion from macroscopic information shown
on the fundamental diagram.

This also suggests that the social impacts from the two congestion
dynamics are different. For moderate flow with a smaller area enclosed,
its impact is also small corresponding to the intercept in Fig. \ref{fig:Elossvsarea}.
For serious congestion, the average delay could be huge, growing with
the area enclosed. 

\subsection{Economic Loss in Simulation}

To further investigate the information hidden in the relation between
the loss area and the economic loss, we simulate the loopy dynamics
by the optimal velocity model. In the simulation, we vary the location
of the accident $x_{\text{acc}}=r\left\Vert x_{2}-x_{1}\right\Vert $
from the upstream end, for $0<r<1$. A vehicle at position $x_{\text{acc}}$
is forced to move with speed $v=0.01v_{\text{max}}$ for a time $\tau$.
We control the scale of the accident by tuning $\tau$. It can also
be interpreted as the road clearing time of an accident in reality.
The congestion time interval counts from the start of the control
until the system recovers to the normal state.

\begin{figure}
\begin{centering}
\includegraphics[scale=0.5]{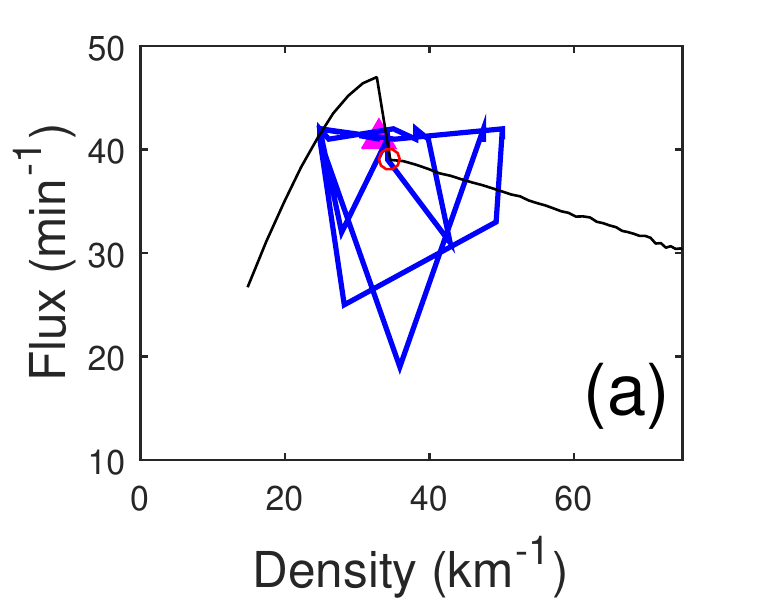}\includegraphics[scale=0.5]{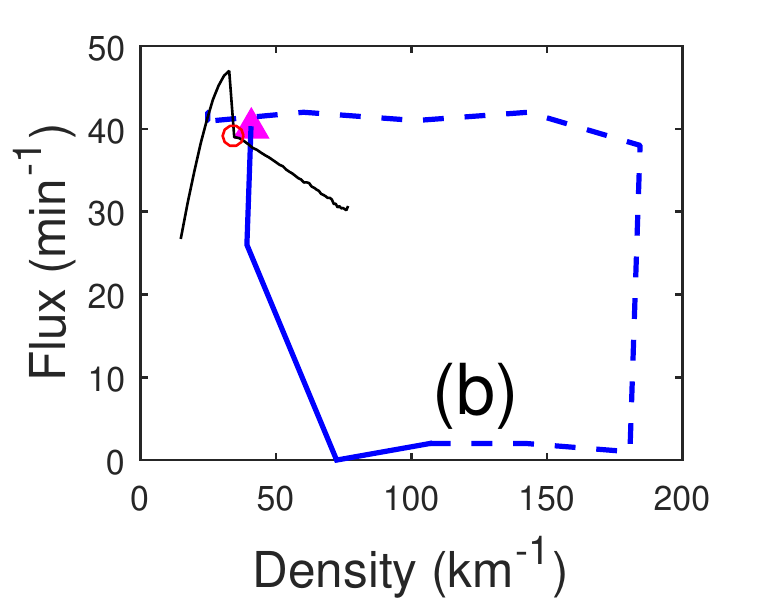}
\par\end{centering}
\caption{\label{fig:twodynamics}(a) Moderate flow and (b) serious congestion
in the simulations. The blue lines indicate the evolutions of the
system during the congestion. The black line is the reference line
as shown in Fig. \ref{fig:OVMfd}. The red circle indicates the state
of the system in simulations. The serious congestion is observed in
the accident condition with $\tau=5\;\text{min}$. The purple triangle
indicates the starting point of the evolution. In Fig. \ref{fig:twodynamics}(b),
two linestyles are employed to illustrate the orientation of the evolution.
The dashed line indicates the evolution of the system in the first
3 min after the accident, while the solid line is the evolution afterwards.
In Fig. \ref{fig:twodynamics}(a), only solid line is used for clarity.
The graphs are drawn in different scale to have a better illustration
of the difference between the dynamics.}

\end{figure}
 The two dynamics classified in real traffic can also be found in
the optimal velocity model. Figures \ref{fig:twodynamics}(a) and
(b) show the moderate flow and serious congestion in simulation, respectively.
Without any accidents introduced, the flux remains constant with small
density fluctuations. By introducing an accident to the system with
$\tau=5\:\text{min}$, $r=0.9$, the serious congestion with an anti-clockwise
loopy dynamics is observed.

\begin{figure}
\begin{centering}
\includegraphics[scale=0.7]{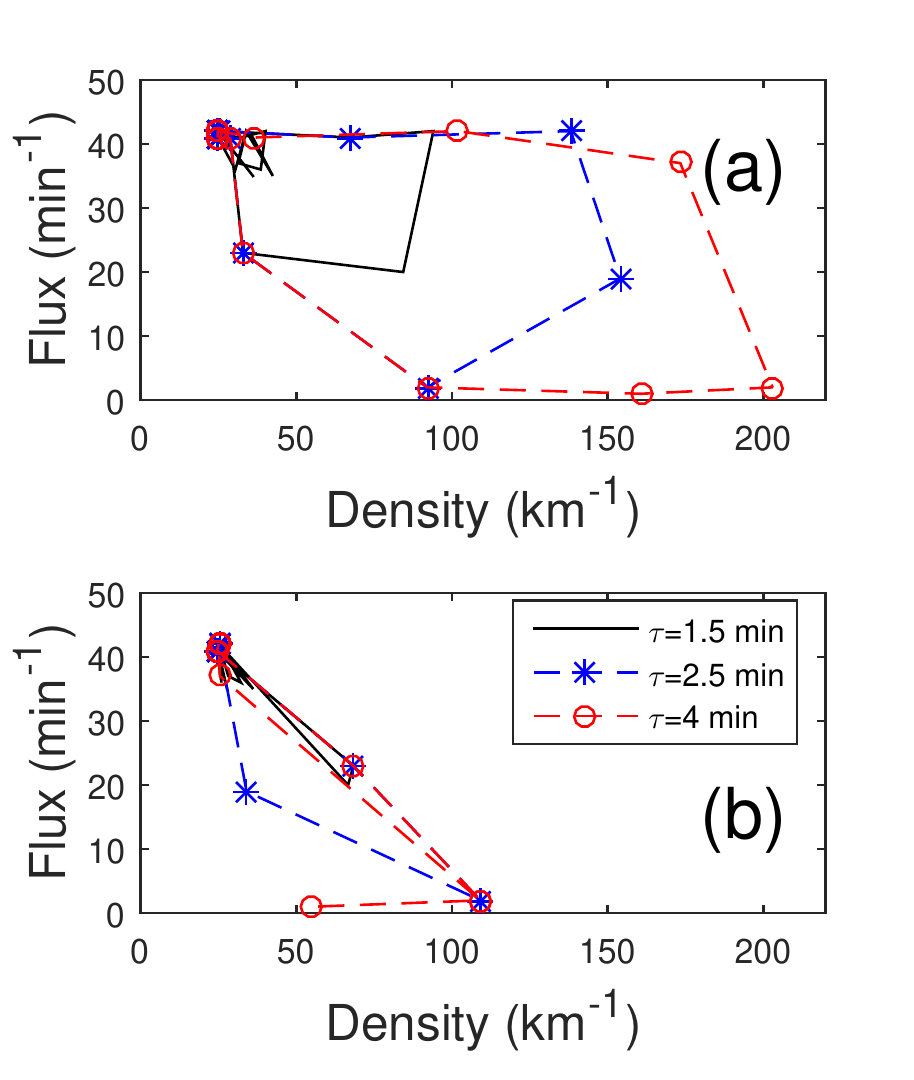}
\par\end{centering}
\caption{\label{fig:accidentcmp}The evolution of the system during accident
with (a) the 2-point measure, and (b) the 1-point measure. The black
solid line, the blue dashed line with stars, and the red dashed line
with circles correspond to different scale of accidents, $\tau=$1.5
min, 2.5 min, and 4 min, respectively.}

\end{figure}
 We compare the 2-point measure and the 1-point measure in describing
the evolution during accidents with different $\tau$. Figure \ref{fig:accidentcmp}(a)
shows a significant increase of the loss area from the 2-point measure
when $\tau$ increases. On the contrary, the area enclosed by trajectories
from the 1-point measure is neither well defined nor correlated with
$\tau$. This shows the advantage of the 2-point measure in describing
behavior in congestion.

\begin{figure}
\begin{centering}
\includegraphics[scale=0.5]{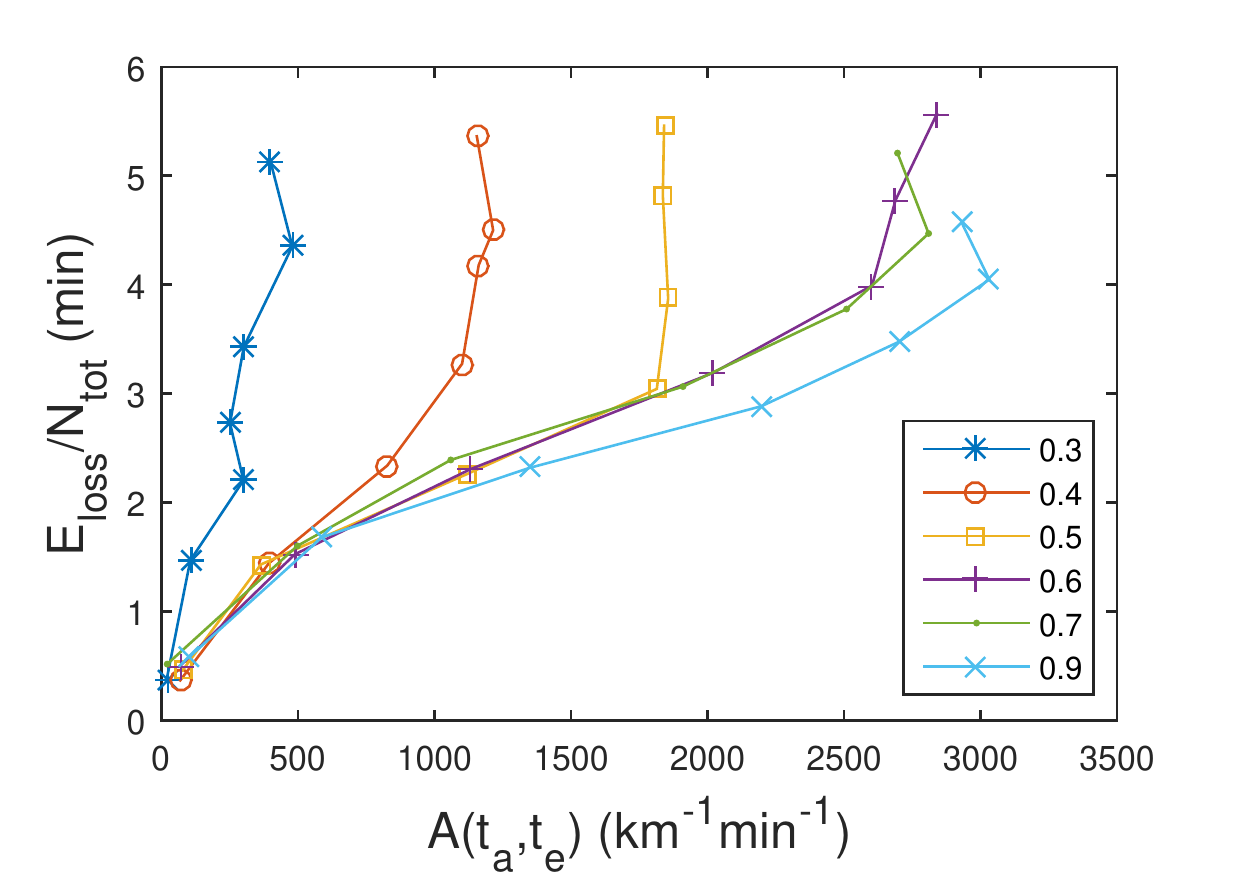}
\par\end{centering}
\caption{\label{fig:OVMrtau}The economic loss per vehicles versus the area
enclosed, for various accident position $x=r\left\Vert x_{2}-x_{1}\right\Vert $
and controlling time $\tau$. Each line corresponds to different values
of $r$, increasing from the leftmost curve, $r=0.3$, to the leftmost
curve, $r=0.9$. Points on the line indicates different values of
$\tau$. With a larger $\tau$, the economic loss per vehicles is
larger.}
\end{figure}
 Figure \ref{fig:OVMrtau} illustrates how the relation between economic
loss $\frac{E_{\text{loss}}}{N_{\text{eff}}}$ and the area enclosed
$A\left(t_{a},t_{e}\right)$ changes with different accident location
$x=r\left\Vert x_{2}-x_{1}\right\Vert $, and road clearing time $\tau$.
For large $r\ge0.6$, $\frac{E_{\text{loss}}}{N_{\text{tot}}}$ increases
linearly with $A\left(t_{a},t_{e}\right)$. This is similar to the
finding from real data. 

Notice that the economic loss saturates in $\tau$ for large $r$,
i.e., the economic loss in the segment does not change with $r$.
This is an artifact due to the imposed periodic boundary condition.
Since the total number of vehicles in the system is fixed, there is
a maximum density in the detecting segment, with the accumulation
$\alpha=N$. The maximum value of $A\left(t_{a},t_{e}\right)$ is
reached when all vehicles in the system are inside the segment during
the accident. This would not happen in real traffic systems, as there
are always vehicles entering the system.

The saturation for small $r$ has a different cause. Because of the
accident, vehicles are queuing up from the accident site $x=r\left\Vert x_{2}-x_{1}\right\Vert $.
If the accident site is closer to the upstream end, the maximum accumulation
of the segment decreases. This results in a smaller $A\left(t_{a},t_{e}\right)$.
Assume that the back propagation speed of congestion is $w$, the
segment reaches the maximum accumulation if $\tau\ge\frac{r\left\Vert x_{2}-x_{1}\right\Vert }{w}$.
In contrast to the saturation for large $r$, this is possible when
an accident is so serious that the road clearing time is too long.
This suggests a possibility in extracting the accident information
(location and road clearing time) from the graph of $\frac{E_{\text{loss}}}{N_{\text{eff}}}$
versus $A\left(t_{a},t_{e}\right)$ through calibration, if we have
both quantities. It is possible to identify the black spot of accidents
on a segment.

\section{Discussion and Conclusion}

To describe the phenomenon of congestion, we have shown that it is
more reliable in using a segment-wise description. The 2-point measure
is more robust to speed fluctuations during congestion, and reflects
the reference density and flux in the simulation of the optimal velocity
model.

Through the consideration of trajectories during congestion, we introduce
the area enclosed by the trajectory as an important quantity in describing
dynamics during congestion. The loss area is shown to have a physical
meaning of the difference between the autocovariance of the outflux
and the temporal covariance between the influx and outflux. 

The loss area constructed by the 2-point measure is much more useful
in the study of the congestion event than its 1-point measure counterpart,
including the identification of the regimes of moderate flow and serious
congestion, and its relation to the economic loss.

Although the information gathered by 1-point measure on two consecutive
detectors is sufficient to obtain the observables in the 2-point measure,
the 2-point measure should not be treated as a simple extension of
the 1-point measure. Equation \eqref{eq:areaenccorr} shows that the
loss area is a product of the congestion time interval with the covariance
between the influx and outflux. The congestion time interval is defined
as the interval when the accumulation or the latency across a segment
is greater than some threshold values. Without the concept of segment
in the 1-point measure, it can only give the covariance between the
influx and outflux, but not the congestion time interval. 

It is also shown that the loss area by the trajectory is proportional
to the economic loss incurred by congestion. Incidentally, this is
reminiscent of hysteresis curves in thermodynamics where the loss
area in the space of stimulus and response represents the energy loss.
With a larger loss area, the average delay incurred is also larger.
This provides a way to estimate the economic loss without the knowledge
of trajectories of every vehicle on the road.

By investigating the relation between the loss area and the flux fluctuation
during congestion, we find that the loss area rises significantly
when the flux fluctuation is greater than a threshold. This supports
the existence of two dynamics during congestion. As illustrated on
the fundamental diagram, one is a loopy evolution, which is triggered
by a sharp drop in flux. The other one is a random fluctuation around
a moderate level of flux independence of density. These two behaviors
are named as ``the serious congestion'' and ``the moderate flow'',
respectively. 

These dynamics should not be treated as evidence of the three phase
traffic theory by Kerner, as they are not related to the behavior
at the downstream front, and much more refined space-time trajectories
of individual vehicles unavailable from our data source are needed
to make further assessment. It is fine to propose two states of congestion,
but tracing the dynamics of the system state in the fundamental diagram
provides a fuller picture of congestion than a time-averaged one.

It is obvious that the serious congestion has a greater impact to
the society, compared to the moderate flow. As congestion is inevitable
due to demand in peak hours, it might be possible to be focus on maintaining
the moderate flow instead. Indeed, flow control by limiting the number
of vehicles entering the system has been implemented in highway systems
\citep{daganzo2007urban}. Our study will be able to quantify the
costs and benefits in optimizing such measures.
\begin{acknowledgments}
We thank the help from Eman Tai, Yulin Xu, and K.T. Siu in managing
the data from Taiwan highway system. This work is supported by the
Research Grants Council of Hong Kong (grant numbers 16322616 and 16306817).
\end{acknowledgments}

\appendix

\section{\label{sec:areacorrl}The Relation of the Area Enclosed with the
Temporal covariance}

Following the definition of the area enclosed in Eq. \eqref{eq:areaenc},
\begin{equation}
A\left(t_{a},t_{e}\right)=\oint\Phi_{\text{out}}\left(t\right)d\rho_{2}\left(t\right),
\end{equation}
where the path of integration is taken to be within the congestion
time interval $\left[t_{a},t_{e}\right]$. When the time interval
is chosen precisely, $\left(\rho_{2}\left(t_{a}\right),\Phi_{\text{out}}\left(t_{a}\right)\right)$
is close to $\left(\rho_{2}\left(t_{e}\right),\Phi_{\text{out}}\left(t_{e}\right)\right)$.
The area enclosed is simplified as
\begin{equation}
A\left(t_{a},t_{e}\right)=\int_{t_{a}}^{t_{e}}\Phi_{\text{out}}\left(t\right)\left(\frac{d}{dt}\rho_{2}\left(t\right)\right)dt,
\end{equation}
From the definition of density in 2-point measure, Eq. \eqref{eq:den2},
and the conservation of flow, Eq. \eqref{eq:acc2},
\begin{equation}
A\left(t_{a},t_{e}\right)=\frac{1}{L}\int_{t_{a}}^{t_{e}}\Phi_{\text{out}}\left(t\right)\left(\Phi_{\text{in}}\left(t\right)-\Phi_{\text{out}}\left(t\right)\right)dt.\label{eq:SimA}
\end{equation}
 By introducing the averaged flux $\overline{\Phi}_{\text{in}},\overline{\Phi}_{\text{out}}$,
\begin{equation}
\overline{\Phi}_{\text{in}}=\frac{1}{\left|t_{e}-t_{a}\right|}\int_{t_{a}}^{t_{e}}\Phi_{\text{in}}\left(t\right)dt,
\end{equation}
\begin{equation}
\overline{\Phi}_{\text{out}}=\frac{1}{\left|t_{e}-t_{a}\right|}\int_{t_{a}}^{t_{e}}\Phi_{\text{out}}\left(t\right)dt,
\end{equation}
we consider the replacement,
\begin{equation}
\Phi_{\text{in}}\left(t\right)=\overline{\Phi}_{\text{in}}+\Phi_{\text{in}}\left(t\right)-\overline{\Phi}_{\text{in}},
\end{equation}
\begin{equation}
\Phi_{\text{out}}\left(t\right)=\overline{\Phi}_{\text{out}}+\Phi_{\text{out}}\left(t\right)-\overline{\Phi}_{\text{out}},
\end{equation}
in Eq. \eqref{eq:SimA}.
\begin{eqnarray}
A\left(t_{a},t_{e}\right)= &  & \frac{\left|t_{e}-t_{a}\right|}{L}\left[\overline{\Phi}_{\text{out}}\overline{\Phi}_{\text{in}}-\overline{\Phi}_{\text{out}}\overline{\Phi}_{\text{out}}\right]+\frac{\left|t_{e}-t_{a}\right|}{L}\label{eq:A2}\\
 &  & \times\left(\text{cov}_{\left[t_{a},t_{e}\right]}\left(\Phi_{\text{in}},\Phi_{\text{out}}\right)-\text{cov}_{\left[t_{a},t_{e}\right]}\left(\Phi_{\text{out}},\Phi_{\text{out}}\right)\right),\nonumber 
\end{eqnarray}
where 
\begin{eqnarray}
\text{cov}_{\left[t_{a},t_{e}\right]}\left(A,B\right)= &  & \frac{1}{\left|t_{e}-t_{a}\right|}\\
 &  & \times\int_{t_{a}}^{t_{e}}\left(A\left(t\right)-\bar{A}\left(t\right)\right)\left(B\left(t\right)-\bar{B}\left(t\right)\right)dt,\nonumber 
\end{eqnarray}
 denotes the covariance of the two time series $A\left(t\right)$
and $B\left(t\right)$ in the time interval $\left[t_{a},t_{e}\right]$.

During $\left[t_{a},t_{e}\right]$, the total number of vehicles entering
(exiting) the segment is $\left|t_{e}-t_{a}\right|\overline{\Phi}_{\text{in}}$
($\left|t_{e}-t_{a}\right|\overline{\Phi}_{\text{out}}$). Since the
system fully recovers from the congestion at $t_{e}$, 
\begin{equation}
\overline{\Phi}_{\text{in}}\approx\overline{\Phi}_{\text{out}}.\label{eq:inoutflux}
\end{equation}
The contribution from the first term in Eq. \eqref{eq:A2} would be
negligible. The area enclosed $A\left(t_{a},t_{e}\right)$ would be
dominated by the covariance term. Hence,
\begin{eqnarray}
A\left(t_{a},t_{e}\right)\approx &  & \frac{\left|t_{e}-t_{a}\right|}{L}\\
 &  & \times\left(\text{cov}_{\left[t_{a},t_{e}\right]}\left(\Phi_{\text{in}},\Phi_{\text{out}}\right)-\text{cov}_{\left[t_{a},t_{e}\right]}\left(\Phi_{\text{out}},\Phi_{\text{out}}\right)\right).\nonumber 
\end{eqnarray}

\section{\label{sec:Approximation-of-the}Approximate Relation between the
Economic Loss and the Area Enclosed}

Following the physical picture of the loopy dynamics during congestion,
we consider an accident happening in the system and congestion propagates
backward to the upstream end. By computing the economic loss and the
area enclosed in this situation, we have an approximate relation between
them.

Suppose an accident happens on a segment of length $L$ and the segment
is completely blocked at the accident site in $\left[t_{a},t_{e}\right]$.
The serious congestion consists of two major stages, the queueing
of vehicles and the recovery to the free flow. We assume that the
traffic conditions in the upstream and downstream segments are similar,
which also agrees with our observations of the time series in the
real data. Then the time taken in each stage is roughly the same,
and can be approximated as $\frac{1}{2}\left|t_{e}-t_{a}\right|$.
For a sufficiently long $\left[t_{a},t_{e}\right]$, the relaxation
time of congestion can be negligible, and hence, every vehicles spent
$\frac{1}{2}\left|t_{e}-t_{a}\right|$ more time on the segment. The
economic loss caused by this congestion is approximated by
\begin{equation}
E_{\text{loss}}=\frac{1}{2}\left|t_{e}-t_{a}\right|N_{\text{tot}},
\end{equation}
\begin{equation}
N_{\text{tot}}=\int_{t_{a}}^{t_{e}}\Phi_{\text{in}}\left(t\right)dt,
\end{equation}
with $N_{\text{tot}}$ being the number of vehicles using the segment
during congestion. We assume that the system is in a steady state
before the accident happens, and hence the flux is uniform over the
whole segment, i.e. $\Phi_{\text{out}}\approx\Phi_{\text{in}}=\Phi_{\text{acc}}$,
where $\Phi_{\text{acc}}$ denotes the flux over the whole system
just before the accident. During this congestion, the system will
evolve through an anticlockwise loop on the fundamental diagram. The
outflux drop from $\Phi_{\text{acc}}$ to $0$ by assumption. 
\begin{equation}
\varDelta\Phi_{\text{out}}=-\Phi_{\text{acc}}.\label{eq:Bfluxacc}
\end{equation}
The accumulation grows according to the number of vehicles entering
the system during the time interval,
\begin{equation}
\varDelta\alpha\approx\Phi_{\text{acc}}\left|t_{e}-t_{a}\right|.\label{eq:Balpha}
\end{equation}
Here, it is assumed that the accident in the segment does not affect
the influx, and the segment upstream of the accident site is not yet
fully occupied during $\left[t_{a},t_{e}\right]$. The area enclosed
during this accident would be
\begin{equation}
A\left(t_{a},t_{e}\right)\approx\left|\varDelta\Phi_{\text{out}}\right|\times\frac{\varDelta\alpha}{L}=\frac{\Phi_{\text{acc}}^{2}}{L}\left|t_{e}-t_{a}\right|.\label{eq:BArea}
\end{equation}
\begin{equation}
\frac{E_{\text{loss}}}{N_{\text{tot}}}\approx\frac{L}{2\Phi_{\text{acc}}^{2}}A\left(t_{a},t_{e}\right).\label{eq:eloss1}
\end{equation}

If we consider the general case that the segment is only partially
blocked by the accident, then the outflux $\Phi_{\text{out}}$ drop
by a fraction $\left(1-f\right)$ to $\left(1-f\right)\Phi_{\text{acc}}$.
$\Phi_{\text{acc}}$ in Eqs. \eqref{eq:Bfluxacc} to \eqref{eq:BArea}
should be replaced by $f\Phi_{\text{acc}}$. On the other hand, it
is noted that under this condition, not all vehicles are delayed by
$\frac{1}{2}\left|t_{e}-t_{a}\right|$. We simplify the situation
by assuming that the vehicles are noninteracting; one possible scenario
for a multi-lane highway is that vehicles along some lanes are blocked
by accidents and are unable to switch lanes while those along other
lanes can travel at normal speed. We then arrive at an effective model
in which vehicles are separated into two classes, the affected vehicles
and the unaffected vehicles. The number of affected vehicles can be
approximated by $fN_{\text{tot}}$. There is negligible delay for
the unaffected vehicles, and hence the economic loss per vehicle should
be
\begin{equation}
\frac{E_{\text{loss}}}{N_{\text{tot}}}\approx\frac{1}{N_{\text{tot}}}\left(fN_{\text{tot}}\left(\frac{1}{2}\left|t_{e}-t_{a}\right|\right)+\left(1-f\right)N_{\text{tot}}\times0\right),
\end{equation}
\begin{equation}
\frac{E_{\text{loss}}}{N_{\text{tot}}}\approx\frac{L}{2f\Phi_{\text{acc}}^{2}}A\left(t_{a},t_{e}\right).
\end{equation}
In fact, $\Phi_{\text{acc}}$ is bounded above by the maximum flux
at free flow, $\Phi_{\text{max}}$ due the physical limit of road
capacity. Hence the area enclosed could give an approximation of the
lower bound of the averaged economic loss $\frac{E_{\text{loss}}}{N_{\text{tot}}}$.
\begin{equation}
\frac{E_{\text{loss}}}{N_{\text{tot}}}\gtrsim\frac{L}{2f\Phi_{\text{max}}^{2}}A\left(t_{a},t_{e}\right).\label{eq:avelossbd}
\end{equation}
It is possible to approximate the number of affected vehicles $N_{\text{tot}}$
by its lower bound, which is the number of vehicles coming in during
congestion, $\Phi\left|t_{e}-t_{a}\right|$. The lower bound of the
total economic loss $E_{\text{loss}}$ would be given by 
\begin{equation}
E_{\text{loss}}\gtrsim\frac{L}{2f\Phi_{\text{max}}}\left|t_{e}-t_{a}\right|A\left(t_{a},t_{e}\right).\label{eq:elossbd}
\end{equation}


\begin{thebibliography}{37}
\providecommand{\natexlab}[1]{#1}
\providecommand{\url}[1]{\texttt{#1}}
\expandafter\ifx\csname urlstyle\endcsname\relax
  \providecommand{\doi}[1]{doi: #1}\else
  \providecommand{\doi}{doi: \begingroup \urlstyle{rm}\Url}\fi

\bibitem[Cole et~al.(2003)Cole, Dodis, and Roughgarden]{cole2003pricing}
Richard Cole, Yevgeniy Dodis, and Tim Roughgarden.
\newblock Pricing network edges for heterogeneous selfish users.
\newblock In \emph{Proceedings of the thirty-fifth annual ACM symposium on
  Theory of computing}, pages 521--530. ACM, 2003.

\bibitem[Cooper(1981)]{cooper1981introduction}
Robert~B Cooper.
\newblock \emph{Introduction to queueing theory}.
\newblock North Holland,, 1981.

\bibitem[Daganzo(1995)]{daganzo1995requiem}
Carlos~F Daganzo.
\newblock Requiem for second-order fluid approximations of traffic flow.
\newblock \emph{Transportation Research Part B: Methodological}, 29\penalty0
  (4):\penalty0 277--286, 1995.

\bibitem[Daganzo(2007)]{daganzo2007urban}
Carlos~F Daganzo.
\newblock Urban gridlock: Macroscopic modeling and mitigation approaches.
\newblock \emph{Transportation Research Part B: Methodological}, 41\penalty0
  (1):\penalty0 49--62, 2007.

\bibitem[Davis(2008)]{davis2008driver}
LC~Davis.
\newblock Driver choice compared to controlled diversion for a freeway double
  on-ramp in the framework of three-phase traffic theory.
\newblock \emph{Physica A: Statistical Mechanics and its Applications},
  387\penalty0 (25):\penalty0 6395--6410, 2008.

\bibitem[Edie(1963)]{edie1963discussion}
Leslie~C Edie.
\newblock \emph{Highway capacity manual}.
\newblock Port of New York Authority, 1963.

\bibitem[Gao et~al.(2007)Gao, Jiang, Hu, Wang, and Wu]{gao2007cellular}
Kun Gao, Rui Jiang, Shou-Xin Hu, Bing-Hong Wang, and Qing-Song Wu.
\newblock Cellular-automaton model with velocity adaptation in the framework of
  kerner three-phase traffic theory.
\newblock \emph{Physical Review E}, 76\penalty0 (2):\penalty0 026105, 2007.

\bibitem[Geroliminis et~al.(2007)Geroliminis, Daganzo,
  et~al.]{geroliminis2007macroscopic}
Nikolas Geroliminis, Carlos~F Daganzo, et~al.
\newblock Macroscopic modeling of traffic in cities.
\newblock In \emph{TRB 86th annual meeting}, number 07-0413, 2007.

\bibitem[Gupta and Dhiman(2015)]{gupta2015phase}
Arvind~Kumar Gupta and Isha Dhiman.
\newblock Phase diagram of a continuum traffic flow model with a static
  bottleneck.
\newblock \emph{Nonlinear Dynamics}, 79\penalty0 (1):\penalty0 663--671, 2015.

\bibitem[Hoogendoorn et~al.(2008)Hoogendoorn, van Lint, and
  Knoop]{hoogendoorn2008macroscopic}
Serge~P Hoogendoorn, Hans van Lint, and Victor~L Knoop.
\newblock Macroscopic modeling framework unifying kinematic wave modeling and
  three-phase traffic theory.
\newblock \emph{Transportation Research Record}, 2088\penalty0 (1):\penalty0
  102--108, 2008.

\bibitem[Jia et~al.(2011)Jia, Li, Chen, Jiang, and Gao]{jia2011cellular}
Bin Jia, Xin-Gang Li, Tao Chen, Rui Jiang, and Zi-You Gao.
\newblock Cellular automaton model with time gap dependent randomisation under
  kerner three phase traffic theory.
\newblock \emph{Transportmetrica}, 7\penalty0 (2):\penalty0 127--140, 2011.

\bibitem[Jin et~al.(2013)Jin, Wang, Jiang, Zhang, and Wang]{jin2013spontaneous}
Cheng-Jie Jin, Wei Wang, Rui Jiang, HM~Zhang, and Hao Wang.
\newblock Spontaneous phase transition from free flow to synchronized flow in
  traffic on a single-lane highway.
\newblock \emph{Physical Review E}, 87\penalty0 (1):\penalty0 012815, 2013.

\bibitem[Kerner(1999{\natexlab{a}})]{kerner1999congested}
Boris Kerner.
\newblock Congested traffic flow: Observations and theory.
\newblock \emph{Transportation Research Record: Journal of the Transportation
  Research Board}, \penalty0 (1678):\penalty0 160--167, 1999{\natexlab{a}}.

\bibitem[Kerner(1998)]{kerner1998experimental}
Boris~S Kerner.
\newblock Experimental features of self-organization in traffic flow.
\newblock \emph{Physical review letters}, 81\penalty0 (17):\penalty0 3797,
  1998.

\bibitem[Kerner(1999{\natexlab{b}})]{kerner1999physics}
Boris~S Kerner.
\newblock The physics of traffic.
\newblock \emph{Physics World}, 12\penalty0 (8):\penalty0 25,
  1999{\natexlab{b}}.

\bibitem[Kerner(2005)]{kerner2005physics}
Boris~S Kerner.
\newblock The physics of traffic: empirical freeway pattern features,
  engineering applications, and theory.
\newblock \emph{Physics Today}, 58\penalty0 (11):\penalty0 54--56, 2005.

\bibitem[Kerner(2016)]{kerner2016failure}
Boris~S Kerner.
\newblock Failure of classical traffic flow theories: stochastic highway
  capacity and automatic driving.
\newblock \emph{Physica A: Statistical Mechanics and its Applications},
  450:\penalty0 700--747, 2016.

\bibitem[Kerner and Rehborn(1996)]{kerner1996experimental}
Boris~S Kerner and Hubert Rehborn.
\newblock Experimental properties of complexity in traffic flow.
\newblock \emph{Physical Review E}, 53\penalty0 (5):\penalty0 R4275, 1996.

\bibitem[Kerner et~al.(2014)Kerner, Klenov, and
  Schreckenberg]{kerner2014probabilistic}
Boris~S Kerner, Sergey~L Klenov, and Michael Schreckenberg.
\newblock Probabilistic physical characteristics of phase transitions at
  highway bottlenecks: incommensurability of three-phase and two-phase
  traffic-flow theories.
\newblock \emph{Physical Review E}, 89\penalty0 (5):\penalty0 052807, 2014.

\bibitem[Kerner(2002)]{kerner2002synchronized}
BS~Kerner.
\newblock Synchronized flow as a new traffic phase and related problems for
  traffic flow modelling.
\newblock \emph{Mathematical and Computer Modelling}, 35\penalty0
  (5-6):\penalty0 481--508, 2002.

\bibitem[Lighthill and Whitham(1955)]{lighthill1955kinematic}
Michael~James Lighthill and Gerald~Beresford Whitham.
\newblock On kinematic waves ii. a theory of traffic flow on long crowded
  roads.
\newblock \emph{Proc. R. Soc. Lond. A}, 229\penalty0 (1178):\penalty0 317--345,
  1955.

\bibitem[McGehee et~al.(2000)McGehee, Mazzae, and Baldwin]{mcgehee2000driver}
Daniel~V McGehee, Elizabeth~N Mazzae, and GH~Scott Baldwin.
\newblock Driver reaction time in crash avoidance research: validation of a
  driving simulator study on a test track.
\newblock In \emph{Proceedings of the human factors and ergonomics society
  annual meeting}, volume~44, pages 3--320. SAGE Publications Sage CA: Los
  Angeles, CA, 2000.

\bibitem[Nagel and Schreckenberg(1992)]{nagel1992cellular}
Kai Nagel and Michael Schreckenberg.
\newblock A cellular automaton model for freeway traffic.
\newblock \emph{Journal de physique I}, 2\penalty0 (12):\penalty0 2221--2229,
  1992.

\bibitem[Nagel et~al.(2003)Nagel, Wagner, and Woesler]{nagel2003still}
Kai Nagel, Peter Wagner, and Richard Woesler.
\newblock Still flowing: Approaches to traffic flow and traffic jam modeling.
\newblock \emph{Operations research}, 51\penalty0 (5):\penalty0 681--710, 2003.

\bibitem[Nagelocd and Rasmussenaf(1994)]{nagelocd1994traffic}
Kai Nagelocd and Steen Rasmussenaf.
\newblock Traffic at the edge of chaos.
\newblock In \emph{Artificial Life IV: Proceedings of the Fourth International
  Workshop on the Synthesis and Simulation of Living Systems}, volume~4, page
  222. MIT Press, 1994.

\bibitem[Newell(1993)]{newell1993simplified}
Gordon~F Newell.
\newblock A simplified theory of kinematic waves in highway traffic, part i:
  General theory.
\newblock \emph{Transportation Research Part B: Methodological}, 27\penalty0
  (4):\penalty0 281--287, 1993.

\bibitem[Ni et~al.(2004)Ni, Leonard, and Hall]{ni2004direct}
Daiheng Ni, John~D Leonard, and Marston Hall.
\newblock Direct methods of determining traffic stream characteristics by
  definition.
\newblock \emph{Transportation Research Record}, 2004.

\bibitem[Orosz et~al.(2009)Orosz, Wilson, Szalai, and
  St{\'e}p{\'a}n]{orosz2009exciting}
G{\'a}bor Orosz, R~Eddie Wilson, R{\'o}bert Szalai, and G{\'a}bor
  St{\'e}p{\'a}n.
\newblock Exciting traffic jams: nonlinear phenomena behind traffic jam
  formation on highways.
\newblock \emph{Physical review E}, 80\penalty0 (4):\penalty0 046205, 2009.

\bibitem[Orosz et~al.(2010)Orosz, Wilson, and St{\'e}p{\'a}n]{orosz2010traffic}
G{\'a}bor Orosz, R~Eddie Wilson, and G{\'a}bor St{\'e}p{\'a}n.
\newblock Traffic jams: dynamics and control, 2010.

\bibitem[Richards(1956)]{richards1956shock}
Paul~I Richards.
\newblock Shock waves on the highway.
\newblock \emph{Operations research}, 4\penalty0 (1):\penalty0 42--51, 1956.

\bibitem[Sch{\"o}nhof and Helbing(2007)]{schonhof2007empirical}
Martin Sch{\"o}nhof and Dirk Helbing.
\newblock Empirical features of congested traffic states and their implications
  for traffic modeling.
\newblock \emph{Transportation Science}, 41\penalty0 (2):\penalty0 135--166,
  2007.

\bibitem[Sch{\"o}nhof and Helbing(2009)]{schonhof2009criticism}
Martin Sch{\"o}nhof and Dirk Helbing.
\newblock Criticism of three-phase traffic theory.
\newblock \emph{Transportation Research Part B: Methodological}, 43\penalty0
  (7):\penalty0 784--797, 2009.

\bibitem[Sugiyama(1999)]{sugiyama1999optimal}
Y{\=u}ki Sugiyama.
\newblock Optimal velocity model for traffic flow.
\newblock \emph{Computer Physics Communications}, 121:\penalty0 399--401, 1999.

\bibitem[Sugiyama et~al.(2008)Sugiyama, Fukui, Kikuchi, Hasebe, Nakayama,
  Nishinari, Tadaki, and Yukawa]{sugiyama2008traffic}
Yuki Sugiyama, Minoru Fukui, Macoto Kikuchi, Katsuya Hasebe, Akihiro Nakayama,
  Katsuhiro Nishinari, Shin-ichi Tadaki, and Satoshi Yukawa.
\newblock Traffic jams without bottlenecks - experimental evidence for the
  physical mechanism of the formation of a jam.
\newblock \emph{New journal of physics}, 10\penalty0 (3):\penalty0 033001,
  2008.

\bibitem[Washington(1985)]{washington1985highway}
D~Washington.
\newblock Highway capacity manual.
\newblock \emph{Special Report}, 209, 1985.

\bibitem[Whitham(2011)]{whitham2011linear}
Gerald~Beresford Whitham.
\newblock \emph{Linear and nonlinear waves}, volume~42.
\newblock John Wiley \& Sons, 2011.

\bibitem[Yang and Wang(2011)]{yang2011managing}
Hai Yang and Xiaolei Wang.
\newblock Managing network mobility with tradable credits.
\newblock \emph{Transportation Research Part B: Methodological}, 45\penalty0
  (3):\penalty0 580--594, 2011.

\end{thebibliography}
\end{document}